\documentclass[format=acmsmall, review=false, screen=true]{acmart}

\usepackage{booktabs} 
\usepackage[caption=false]{subfig} 
\usepackage{tikz}
\usepackage{pgfplots}
\pgfplotsset{compat=1.14, select coords between index/.style 2 args={
    x filter/.code={
        \ifnum\coordindex<#1\fi
        \ifnum\coordindex>#2\fi
    }
}}

\usepackage[ruled]{algorithm2e} 

\SetAlFnt{\small}
\SetAlCapFnt{\small}
\SetAlCapNameFnt{\small}
\SetAlCapHSkip{0pt}
\IncMargin{-\parindent}

\begin{document}
\title[Runtime Mitigation of Packet Drop Attacks in Fault-tolerant NoCs]{Runtime Mitigation of Packet Drop Attacks in Fault-tolerant Networks-on-Chip}  

\author{N. Prasad}
\email{nprasad@ece.iitkgp.ernet.in}
\author{Navonil Chatterjee}
\email{navonil@iitkgp.ac.in}
\author{Santanu Chattopadhyay} 
\email{santanu@ece.iitkgp.ernet.in}
\author{Indrajit Chakrabarti}
\email{indrajit@ece.iitkgp.ernet.in}
\affiliation{%
  \institution{Indian Institute of Technology Kharagpur}
  \department{Department of Electronics and Electrical Communication Engineering}
  \city{Kharagpur}
  \state{WB}
  \postcode{721302}
  \country{India}
}

\begin{abstract}
Fault-tolerant routing (FTR) in Networks-on-Chip (NoCs) has become a common practice to sustain the performance of multi-core systems with an increasing number of faults on a chip. On the other hand, usage of third-party intellectual property blocks has made security a primary concern in modern day designs. This article presents a mechanism to mitigate a denial-of-service attack, namely packet drop attack, which may arise due to the hardware Trojans (HTs) in NoCs that adopt FTR algorithms. HTs, associated with external kill switches, are conditionally triggered to enable the attack scenario. Security modules, such as authentication unit, buffer shuffler, and control unit, have been proposed to thwart the attack in runtime and restore secure packet flow in the NoC. These units work together as a shield to safeguard the packets from proceeding towards the output ports with faulty links. Synthesis results show that the proposed secure FT router, when compared with a baseline FT router, has area and power overheads of at most 4.04\% and 0.90\%, respectively. Performance evaluation shows that SeFaR has acceptable overheads in the execution time, energy consumption, average packet latency, and power-latency product metrics when compared with a baseline FT router while running real benchmarks, as well as synthetic traffic. Further, a possible design of a comprehensive secure router has been presented with a view to addressing and mitigating multiple attacks that can arise in the NoC routers.
\end{abstract}


\maketitle


\section{Introduction}
\label{sec:introduction}
Network-on-Chip (NoC) has emerged as a promising and scalable medium for interconnecting various cores in a multi-core system \cite{noc}. However, with aggressive technology scaling, as well as an increase in the transistor density, multi-core systems are subjected to faults that may occur either during the manufacturing time or runtime. NoCs, being part of such multi-core systems, are also susceptible to these faults. Several fault-tolerant techniques have been developed over the years to overcome the consequences that may arise due to the occurrence of faults in NoCs \cite{csur2013,csur2016}. As mentioned in \cite{csur2013}, the adopted fault-tolerant techniques differ with different layers of the NoC-based multi-core system, such as transport, network, and data-link. Further, in the network layer, which consists of components such as routers and links, fault-tolerant mechanisms are realized through spatial and temporal redundancy techniques. These techniques include redundant modules inside the routers, as well as fault-tolerant routing algorithms in the NoC.

On the other hand, with the ever-increasing complexity of multi-core systems to process diverse tasks, the role of third-party intellectual property (3PIP) cores is increasingly becoming prominent in such systems. With so many 3PIP cores at hand during the system integration, there exists a barrier of security and trust among the 3PIP vendors that tempts the adversaries to adopt unfair means to disrupt the functionality of the other 3PIP cores. One way of achieving such an unsought disruption is through the insertion of hardware Trojans (HTs) into the chip, by an untrusted foundry or design house that include untrusted people, design tools, or components. Consequences of such an uninvited insertion of the HTs include the undesired functional behavior of an integrated circuit (IC) and provision of covert channels or backdoor that can leak sensitive information \cite{HTpieee}. Though several detection techniques, to detect HTs, have been proposed in the literature \cite{HTiscas}, they may not be advantageous for all kinds of HTs present on a chip. Thus, the proactive security measures at the hardware level are increasingly being considered, which can improve the detection capability, as well as mitigate the consequences that arise due to the HTs.

As the NoC forms the backbone of the interconnection architecture of a multi-core system, there is an immense need in securing its hardware resources, as well as to secure the packet flow in it. Hardware security in NoCs has been a vibrant research area for over a decade \cite{nocsec1,nocsec2}. Traditionally, research in secure NoCs has focused on the areas like secure memory access, memory protection, and secure packet flow, in NoC-based multi-core systems \cite{tc,Kapoor2013,iav}. Of late, research in several attacks on NoC hardware, such as attacks due to HTs on routers \cite{FortNoCs,cssp15,JayashankaraShridevi2017,ISCAS17,Frey201715} and links \cite{DATE16,Boraten2017}, as well as timing side-channel attacks \cite{esl}, are being investigated. All the works mentioned in the literature assume that the NoC is a fault-free one. Security aware design of FTNoCs has not been explored to date. With the introduction of faults, many possibilities arise for the adversaries to take advantage of such scenarios to raise new attacks in the NoC. This article proposes one such attack called packet drop attack in the context of FTNoCs and presents a mitigation mechanism to thwart such attacks effectively.

Packet drop attack is similar to a well-known denial-of-service (DoS) attack in a network router, namely blackhole attack, where the router, instead of relaying the packets to proper output ports, discards them \cite{bha04,Tseng2011}. Similar attack scenario can be achieved in NoCs with faulty links, where a router with faulty links, instead of sending packets to ports with healthy links, forwards the packets to ports with faulty links. This scenario is same as dropping the packets by a router, as it is unable to relay them to the next router. Since this case would arise in NoCs in the presence of faulty links, existing mitigation mechanisms for packet drop attacks in the case of wireless or mobile networks may not be advantageous for NoCs. Existing methods to mitigate the packet drop attack in wireless ad hoc and mobile networks, such as routing recovery, time-based threshold, and Bayesian detection, cost a lot concerning area and energy consumption when considered for NoC routers \cite{SURV11,Tseng2011}. Thus, to mitigate the packet drop attacks in FTNoCs, we endow the routers with security modules, such as authentication unit (AU), buffer shuffler (BS), and control unit (CU). The proposed logic-based mitigation mechanism is advantageous when compared to the above-mentioned mechanisms while considering the hardware and timing overheads into account.

The main contributions of the present article are as follows. A mechanism to raise packet drop attacks in the context of fault-tolerant NoCs has been presented. In the current study, we assume that fault-tolerant routing is the fault-tolerant method adopted in the NoC. A possible threat model, as well as design of hardware Trojan that has the potential to raise the packet drop attacks, has been proposed. A secure fault-tolerant NoC router, referred to as SeFaR, has been presented to thwart the packet drop attacks at runtime. Experimental analysis for establishing the impact of the proposed attack, as well as for analyzing the performance of the system in the presence of the proposed secure router, has been presented. A possible design of a comprehensive secure router has been presented, which can mitigate multiple attacks that can arise in the NoC routers.

The rest of the article is organized as follows. Section \ref{sec:relwork} discusses the related work. Section \ref{sec:motivation} presents motivation behind the packet drop attack, proposes the threat model, and discusses its relevance. Section \ref{sec:baseline} briefly describes the architecture of the baseline fault-tolerant router. Section \ref{sec:proposed} describes the security modules including the authentication unit, buffer shuffler, and the control unit. Section \ref{sec:results} presents the performance analysis of the mitigation mechanisms. Section \ref{sec:comprouter} presents the architecture of a comprehensive secure router, which can mitigate multiple attack scenarios that arise within the NoC routers. Section \ref{sec:conclusions} concludes the article.

\section{Related Work}
\label{sec:relwork}

    This Section articles out the related work that has been carried out in the areas of fault-tolerant routing, as well as hardware attacks on the NoC architecture, particularly within routers and links.

\subsection{Fault-tolerant Routing in NoCs}
\label{sub:ftrouting}
    Several fault-tolerant techniques have been extensively reviewed in \cite{csur2016}, which also include fault-tolerant routing algorithms for the NoCs. Majority of the fault-tolerant routing strategies consider distributed routing, whereas very few strategies consider source routing. Next, we review few other fault-tolerant routing algorithms for the NoCs that have been proposed recently.
    
    Liu \textit{et al.} \cite{TCAD15} have proposed two novel adaptive routing algorithms, namely coarse and fine-grained look-ahead routing algorithms, to enhance the fault-tolerant capabilities of 2D mesh/torus NoC system. These strategies use fault flag codes from the neighboring nodes to obtain the status or conditions of real-time traffic in a NoC region, then calculate the path weights and choose the route to forward the packets. Chen \textit{et al.} \cite{TPDS16} have proposed a Path-Diversity-Aware Fault-Tolerant Routing (PDA-FTR) algorithm, which simultaneously considers path diversity information and buffer information. This strategy achieves fault-resilient packet delivery and traffic balancing in the NoC.
    
    Xie \textit{et al.} \cite{TINF17} have presented a Preferable Mad-y (PMad-y) turn model and Low-cost Adaptive and Fault-tolerant Routing (LAFR) method that use one and two virtual channels along the X and Y dimensions for 2D mesh NoC. Applying PMad-y rules and using the link status of neighbor routers within 2-hops, LAFR can tolerate multiple faulty links and routers in more complicated faulty situations and impose the reliability of network without losing the performance of the network. Zhao \textit{et al.} \cite{TC17} have proposed a path-counter method, which labels every node that is helpless to make up for a fault-tolerant minimal path with low time complexity. This strategy can support arbitrary fault distribution, check the existence of fault-tolerant minimal paths, and not sacrifice any available fault-tolerant minimal paths.
    
    Though the FT routing algorithms mentioned above adopt different techniques in successfully transmitting the packets to their destinations if there exists a path, a common metric used in many of these routing algorithms is the knowledge of the status of the links at different capacities in the NoC. Another feature of these routing algorithms is that despite with an increase in the number of faulty links in the NoC, the throughput degradation, as well as the increase in the average packet latency, is kept limited to a considerable value.

\subsection{HT Attacks on NoCs}
\label{sub:htattack}

\begin{figure}[!ht]
    \centering
    \includegraphics[width = \textwidth]{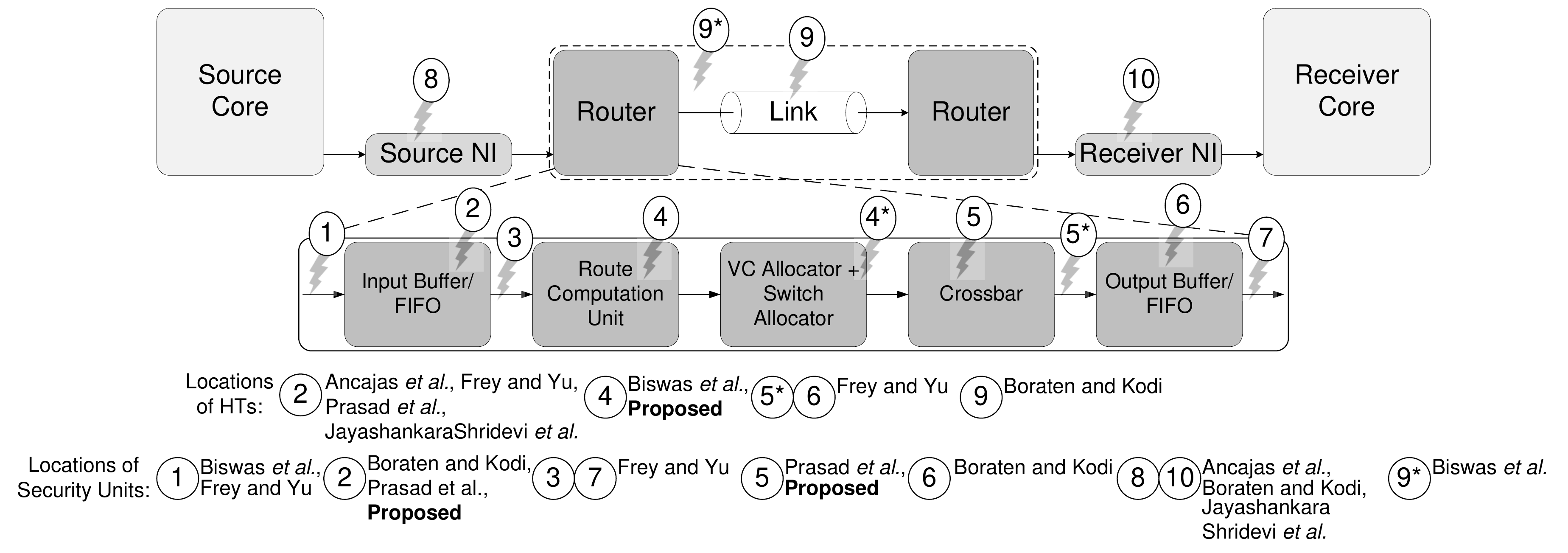}
    \caption{Locations considered in inserting HTs as well as employing security units in NoC based systems.}
    \label{fig:survey}
\end{figure}

    Fig. \ref{fig:survey} shows the locations considered in inserting HTs as well as employing security units (encircled numbers) in NoC-based systems. The following survey primarily considers the attacks on NoC hardware due to HTs.

    Ancajas \textit{et al.} \cite{FortNoCs} have demonstrated a range of security attacks, which include information leaking attacks, that could arise in a compromised NoC (C-NoC) that has an accomplice software component. A series of techniques have been proposed to counter the demonstrated attacks by hardening security on systems with a C-NoC, which consists of data scrambling, packet certification, and node obfuscation. The attack scenario and security measures addressed mainly concentrate on data duplication. No control over the packet flow has been enforced. Biswas \textit{et al.} \cite{cssp15} have investigated a new attack scenario, namely router attack, targeted towards routing tables in routers. These attacks include unauthorized access attack and misrouting attack. The authors have proposed several monitoring-based countermeasures to thwart these attacks. However, this article does not consider attacks on logic-based route computation units.
    
    Boraten and Kodi \cite{DATE16} have proposed a packet validation technique, namely \textit{packet-security} (P-Sec), merged with two robust error detection schemes, to protect C-NoCs from fault injection side-channel attacks and covert HT communication. P-Sec is a security measure incorporated at NI, which primarily deals with packet encryption and decryption. However, the technique has no control over attacks that arise within a router. They \cite{Boraten2017} have also proposed a target-activated sequential payload (TASP) HT model that injects faults into the packets, by inspecting them. The faults injected by the HT trigger the error correction code schemes employed for re-transmitting the packets unnecessarily, which could create network congestion and even deadlocks. To circumvent these threats, the authors have proposed a heuristic threat detection model to classify faults and to discover the HTs within compromised links. In addition to this model, several switch-to-switch link obfuscation methods have been proposed to utilize the compromised links instead of rerouting the packets. Though the work covers transient and permanent fault scenarios associated with links in detecting HTs within links, no measure has been considered for threats that arise within a router.
    
    Frey and Yu \cite{Frey201715} have proposed a collaborative dynamic permutation and flit integrity check method to mitigate the consequences caused by the HTs located in the NoC routers. Though the mentioned security hardening mechanism efficiently thwarts the specific role of the HTs, it has no impact on attacks arising in the locations of a router, other than buffers. Prasad \textit{et al.} \cite{ISCAS17} have proposed a DoS attack, namely illegal packet request attack (IPRA), and have addressed the measures to mitigate the same. The IPRAs are raised by the HTs located within routers, which are triggered when a core attached to the router goes idle. A security unit has been proposed to detect these attacks and mitigate the consequent loss by guiding the control units of the corresponding buffers to either isolate or mask the attacked buffers at runtime. Though the work considers the location of the HTs inside routers, the NoC is assumed as a fault-free one. JayashankaraShridevi \textit{et al.} \cite{JayashankaraShridevi2017} have proposed a runtime latency auditor to counter the threat posed by a rogue NoC (rNoC) that can selectively disrupt the perceived availability of on-chip resources. The proposed mechanism to thwart the bandwidth denial attack enables an MPSoC integrator to monitor the trustworthiness of the deployed NoC throughout the chip lifetime. However, the presented mechanism does not consider the HT attacks arising in the control module of a router, such as a routing unit. A summary of the HT mitigation mechanisms in NoC considered so far has been mentioned in Table \ref{tab:HTmitcomp}.

\begin{table}[!ht]
\centering
\caption{Comparison of HT Mitigation Mechanisms in NoCs}
\label{tab:HTmitcomp}
\resizebox{\textwidth}{!}{
\begin{tabular}{|c|c|c|c|c|}
\hline
Work & \begin{tabular}[c]{@{}c@{}}Trojan \\Location \\in NoC\end{tabular} & Attack(s) Considered & \begin{tabular}[c]{@{}c@{}}Fault-tolerant\\ Router\end{tabular} & \begin{tabular}[c]{@{}c@{}}Protection Mechanism\\ Location\end{tabular} \\ \hline
Ancajas \textit{et al.} \cite{FortNoCs} & Router & Information leaking & No & Network Interface \\ \hline
Biswas \textit{et al.} \cite{cssp15} & Router & \begin{tabular}[c]{@{}c@{}}Unauthorized access,\\ mis-routing\end{tabular} & No & NoC \\ \hline
Boraten and Kodi \cite{DATE16} & Link & \begin{tabular}[c]{@{}c@{}}Fault injection side-channels,\\ covert HT communication\end{tabular} & No & Network Interface \\ \hline
Frey and Yu \cite{Frey201715} & Router & Packet corruption & No & NoC/Router \\ \hline
Prasad \textit{et al.} \cite{ISCAS17} & Router & Illegal packet requests & No & NoC/Router \\ \hline
JayashankaraShridevi \textit{et al.} \cite{JayashankaraShridevi2017} & Router & Bandwidth denial & No & SoC Firmware \\ \hline
Boraten and Kodi \cite{Boraten2017} & Link & fault injection side channels & Yes & NoC/Router \\ \hline
\textbf{Present Work} & \textbf{Router} & \textbf{Packet drop} & \textbf{Yes} & \textbf{NoC/Router} \\ \hline
\end{tabular}}
\end{table}

Many of the works mentioned above take into account attacks on NoC within routers and links, assuming that the NoC is fault free. However, in the presence of faults, mainly permanent link faults, there is vast scope for attackers to attack routers, while taking advantage of the knowledge of faulty links. Several strategies exist in the literature that subsume the operation of NoC in the presence of faulty links \cite{csur2013,csur2016}. Although identification of faulty links has been considered in \cite{Boraten2017}, no security mechanism has been proposed thus far for attacks arising within the routers that are adjacent to the faulty links. One of such attacks may well be created with the motivation of forwarding packets towards the output ports associated with the faulty links. This work has addressed one such attack situation and has proposed a secure router to circumvent those attacks.

\section{Baseline System Model}
\label{sec:baseline}
    This section describes the baseline system model by detailing the architectural and functional properties of the NoC-based MPSoC, which has been considered for the present study.
    
    An MPSoC in the present study is a tile-based system, where each tile consists of a processing element (PE) or core, a set of level-one private instruction and data caches, a memory controller, and a router along with the network interface. A shared level-two cache has also been considered to be present in the system. The communication infrastructure that interconnects all the tiles in the system has been realized using a two-dimensional mesh NoC. Traditional input-queue based buffered routers have been considered, which adopt round-robin arbitration mechanism for switch allocation and logic-based fault-tolerant distributing routing units. An application in the MPSoC has been assumed to be mapped onto multiple cores using the mapping algorithm mentioned in \cite{Sahu}. The nature of the applications considered is generic. Also, the kind of the MPSoC can be either homogeneous or heterogeneous. The communication between the cores is deemed to happen via the NoC, through network interface units and routers in the network.

\begin{figure}[!ht]
\centering
\subfloat[]{\label{fig:rwt}\includegraphics[width=0.25\textwidth]{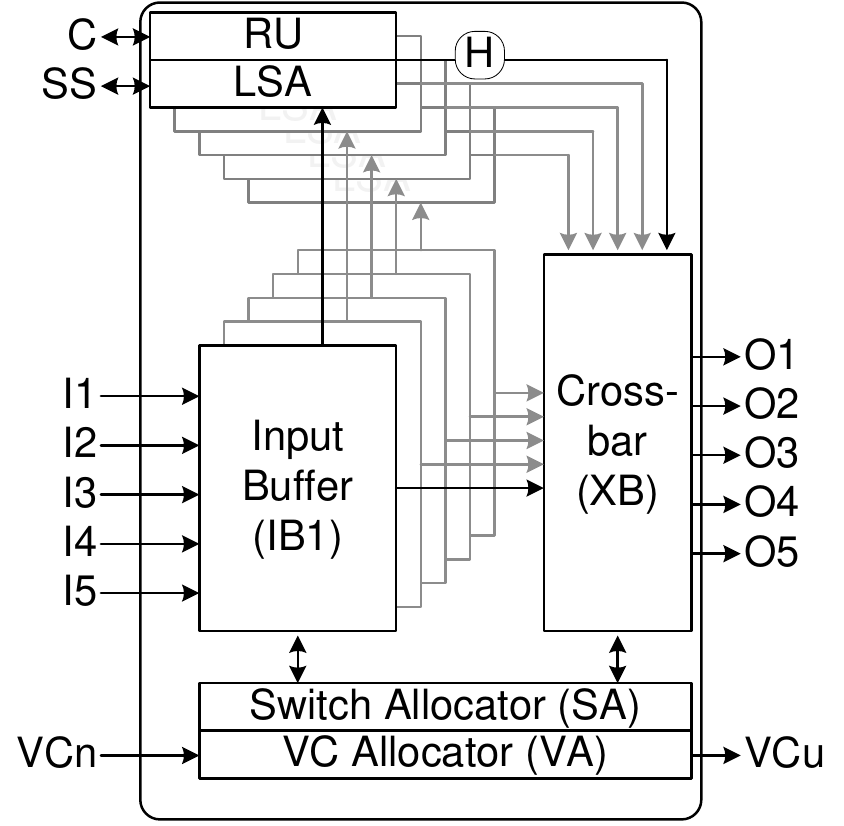}}
\hspace{0.25\textwidth}
\subfloat[]{\label{fig:nwfl}\includegraphics[width=0.225\textwidth]{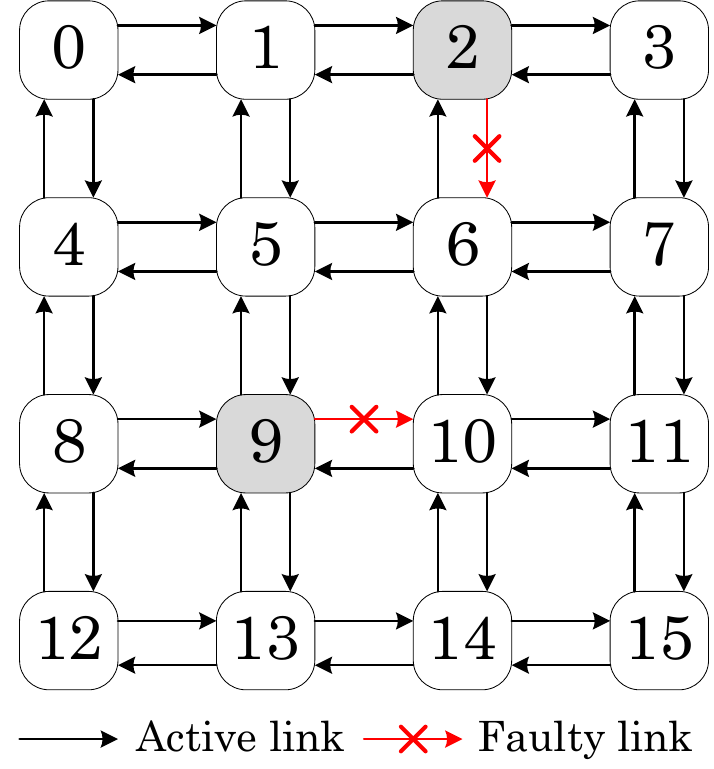}}
\caption{(a) Microarchitecture of a fault-tolerant NoC router with a hardware Trojan. (b) A 4$\times$4 NoC with faulty links.}
\end{figure}

    Fig. \ref{fig:rwt} shows a baseline FT router microarchitecture, which consists of a link status analyzer (LSA) with each routing unit (RU), for forwarding packets to fault-free output ports, in the presence of faulty links. The remaining inputs and outputs of the router are as follows. I1, $\ldots$, I5 and O1, $\ldots$, O5 are the data input and output (IO) ports. Credits (C), which record the occupancy of buffers of adjacent routers, deal with buffer and virtual channel management, and status signals (SS) record the health of the links associated with the ports of the router. All routers in the NoC are assumed to have five-stage pipeline: input buffer and route computation, virtual channel allocation, switch allocation, switch traversal, and link traversal. Regarding the fault-tolerant routing algorithm, we assume that the routing algorithm maintains the status of all the links that are associated with the current router. A current router, concerning a packet, is defined as the router in which the packet currently resides.

    The routing algorithm considered in the baseline fault-tolerant router is similar to the one mentioned in \cite{TCAD15}, where each router maintains a register, named as LSR, in which the status of the links connected to the router is stored. In the current analysis, the state of those links, which are up to two hops away from the current router, has been considered.

\section{Motivation and Threat Relevance}
\label{sec:motivation}
    This section presents the motivation for considering the packet drop attack and describes the threat relevance in the context of NoCs with faulty links.

\subsection{Motivation and Attack Scenario}
\label{sub:motivation}
    Analysis of works surveyed in \cite{csur2013,csur2016}, as well as the ones mentioned in Section \ref{sub:ftrouting}, shows that the measures taken to forward packets in a NoC with faulty links are mostly enforced either at the RU or input buffers. However, the intended inclusion of HTs makes even these units vulnerable from the security perspective. Thus, any attack within the FT router created after the RC pipeline stage goes undiscovered. This drawback motivates the attacker to choose an appropriate attack point to disrupt the functionality of the NoC endowed with FT routers.
    
    
    Consider a 4$\times$4 mesh based two-dimensional NoC as shown in Fig. \ref{fig:nwfl}. The links in black are healthy links, and those in red are faulty. Faults associated with the links are considered to be permanent faults. The mechanism given in \cite{Boraten2017} for identifying permanent faults can be applied here as well to detect the faults. Further, the current study considers that the NoC has not been divided into secure and non-secure zones. Consider routers nine and ten in Fig. \ref{fig:nwfl}. The link connecting from router nine to router ten is a faulty one. Once this link becomes faulty, router nine informs its neighbors regarding the faulty link and sends the updated link status information to them. Consider a case where a packet needs to travel from router 8 to router 11, via routers 9 and 10. Had the link that connects from router 9 to router 10 been healthy, the packet would have traveled through router 9. Since the link is faulty in this case, router 8 will now have the updated link status and the packet from router 8 to router 11 will find another path for its traversal, for instance through the routers 8, 4, 5, 6, 10, and 11. This path avoids router 9 in the network. Consider another scenario where another packet needs to travel from router 8 to router 5, with the path through the routers 8, 9, and 5, being the primary path and the path through the routers 8, 4, and 5, being a secondary path. Though router 8 knows the information of the faulty link at the east output port of router 9, it would still send the packet to router 5 through router 9, as the path connecting routers 8, 9, and 5, does not have any faulty link. Thus, this path does not avoid router 9. Assume now that there is an HT present in router nine at the routing unit associated with its left input port, with a motivation to drop the packets that arrive at this port, by forwarding them to the east output port, which has the faulty link. Though the packets from router 8 to router 10 or router 11 avoid router 9, those other packets, which have to take a turn at router 8, such as the ones from router 8 to router 5, are dropped at router 9, because of the presence of the HT in router 9.
    
    Attack situations as mentioned above, though not common, can be easily created in NoCs, since most of the FTR also consider network congestion information to route the packets in the network \cite{Behrouz2014} adaptively. Further, though there exist techniques, such as packet retransmission, to reduce the loss of packets in traditional FT NoCs, they will not be successful in a scenario as mentioned above. The reason for such techniques for not being successful is because even if the packets are retransmitted, they would still follow the same path, which makes them inevitable from getting lost again. This kind of packet dropping can be done by manipulating the port select signals of the XB, which come from the RU. Had there not been any attack in the scenario mentioned above, and with the link associated with the east port of router nine is faulty, the RU associated with the left input port of router 9 would have adapted its routing algorithm such that no packet would go towards its east output port. Owing to the FT operation of a router, if an attack is created after the RU, it is always possible to forward the packets to the east output port, which has the faulty link. This kind of misrouting and its subsequent packet dropping process ends up in data loss and can be disruptive if deployed in critical systems.

\subsection{Threat Model}
\label{sub:htmodel}
An attacker in the current study can be either a rogue employee who knows a part of the design, a computer-aided design (CAD) tool, an untrusted foundry, or any combination of the above \cite{HTpieee,Frey201715}. The attacker can infect HTs into the design only during the design time. Once the chip is fabricated, the attacker cannot further modify the functionality of the design. Fig. \ref{fig:rwt} shows an attack point of the proposed attack in a router, which is marked as H.

An attacker can infect the system in one of the following ways. A rogue employee of the design house can leak the design information to the foundry, which can replace the correct masks with the ones infected with the HTs \cite{Frey201715}. Otherwise, a CAD tool can hide particular details of the design from the designer while generating the layout of the design. Similarly, an untrusted foundry may reverse engineer the obtained masks to infect the HTs. However, for the present study, we assume that the addition of the security modules to the design is done at the last phase of the design, similar to \cite{Boraten2017}.

Similar to all HTs, the proposed HT is assumed to go undetected either during the design verification phase or the chip testing phase. An HT in the proposed design remains dormant in a router until one of the links associated with the output ports of the router becomes faulty. Once a link associated with an infected router becomes faulty, the HT wakes up and waits until the trigger condition is met to raise the packet drop attack. Once the HT is triggered, it starts to divert the packets in the router to the output ports with faulty links, which is the threat in the case of a packet drop attack. However, unlike in packet drop attacks of other networks, where multiple malicious nodes cooperate, we assume that the attack scope of the proposed HTs is limited to those routers in which they are injected.

An infected router in the proposed design is a baseline fault-tolerant router with an HT. However, a difference between an infected and a non-infected router is that a non-infected router always tries to divert the packets towards the output ports that do not have faulty links. Whereas, an infected router is a router that always tries to divert the packets at an input port towards the output ports that have faulty links.

Since an attacker does not have provision to modify the design once fabricated, placement of HTs in the NoC is essential to increase the chance of triggering the HTs, as well as making the attack a significant one. In the present design, selection of routers in the NoC to place the HTs has been made considering the traffic observed while running real benchmark circuits on a NoC-based MPSoC platform. Further details on the same have been mentioned in Section \ref{sub:attacksig}.


\subsection{Attack Significance}
\label{sub:attacksig}

The attack scenario to raise the packet-drop attacks, as mentioned in Section 4.1, requires a NoC to have at lease one faulty link. The reason to have one or more faulty links is that of the specific behavior of the proposed HT, which remains dormant till a link in the NoC becomes faulty. This specific behavior of the proposed HT limits its significance to create a potential attack scenario. However, with the knowledge of the design footprint of the proposed HT, which will be mentioned in Section 4.4, one may use a combination of the proposed HT with any other HT to increase the scope of the attack scenario in the NoC, and still hide the HTs during the design verification or chip testing phase, to carry out the desired attack.

On the other hand, a NoC architecture can be considered as a truly fault-tolerant one, only if the fault tolerance mechanism addresses all its hardware resources, such as links and routers. Several such architectures, as reviewed in \cite{csur2013,csur2016}, either focus on links alone or focus on both routers and links, to endow the fault tolerance property to the NoC. Thus the scope of the proposed HT, which requires a NoC to be embedded with a fault-tolerant routing mechanism, can be extended to many fault-tolerant NoC architectures mentioned in \cite{csur2013,csur2016}.

Since there exist many RUs in a router, each associated with an input port, any RU can be chosen for inserting the HT. However, selecting a particular port for inserting the HTs is essential, as different ports in different routers have different probabilities of packet transmission. For instance, inserting the HTs at the RU associated with the local input port of a router is more efficient than placing them at other RUs, if the data transmission at the local port is higher than that of other ports. Further, due to the look-ahead routing mechanism adopted in some FT NoCs \cite{lar}, where the packets may not reach the routers that are associated with the faulty links, the selection of routers, and corresponding ports, becomes more critical to increase the severity of the attack.

\begin{table}[!ht]
\centering
\caption{Parameters to Carry out Full System Simulation}
\label{tab:sniperparam}
\resizebox{220pt}{!}{
\begin{tabular}{|l|r|}
\hline
Parameter & Value \\ \hline
System architecture & Xeon X5550 Gainestown \\ \hline
Core frequency & 2.66 GHz \\ \hline
Number of cores & 64 \\ \hline
\multicolumn{2}{|l|}{\textbf{Shared cache}} \\ \hline
Data access time & 30 cycles \\ \hline
Tags access time & 10 cycles \\ \hline
Cache size & 8 MB \\ \hline
Cache block size & 64 KB \\ \hline
\multicolumn{2}{|l|}{\textbf{DRAM}} \\ \hline
Latency & 45 cycles \\ \hline
\multicolumn{2}{|l|}{\textbf{Network}} \\ \hline
Link bandwidth & 128 bits \\ \hline
Hop latency & 1 \\ \hline
dimensions & 2 \\ \hline
Concentration & 1 \\ \hline
\end{tabular}}
\end{table}

\begin{figure}[!ht]
\centering
\begin{tikzpicture}
\begin{axis}[
    width = \textwidth,
    height = 0.35\textwidth,
    xmin = 0,
    xmax = 63,
    xlabel = Router Index,
    ylabel = Link Utilization (\%)
]
    \addplot [draw=black, mark=*, select coords between index={0}{63}] table {blackscholes_LUR_ForPlots.txt};
    \addlegendentry{North}
    \addplot [draw=red, mark=star, select coords between index={64}{127}] table {blackscholes_LUR_ForPlots.txt};
    \addlegendentry{South}
    \addplot [draw=blue, dashed, mark=o, select coords between index={128}{191}] table {blackscholes_LUR_ForPlots.txt};
    \addlegendentry{West}
    \addplot [draw=green, mark=oplus, select coords between index={192}{255}] table {blackscholes_LUR_ForPlots.txt};
    \addlegendentry{East}
\end{axis}
\end{tikzpicture}
\caption{Link utilization of directional links in an 8$\times$8 mesh NoC when running \textit{blackscholes} benchmark.}
\label{fig:linkutilblack}
\end{figure}
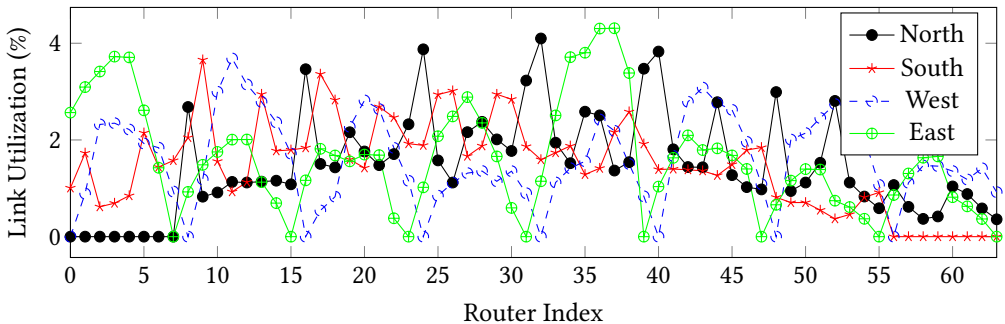

    To establish the significance of the packet drop attack in the present context, traffic distributions of PARSEC \cite{parsec} benchmark applications have been obtained in the forms of traces, link utilization ratios, router utilization ratios, as well as inter-core communication volumes. SniperSim \cite{sniper} full-system simulator has been used to obtain all of the information mentioned above, with the parameters mentioned in Table \ref{tab:sniperparam}. Fig. \ref{fig:linkutilblack} shows the link utilization ratios of the directional links of routers while running the \textit{blackscholes} application in an 8$\times$8 mesh NoC. These ratios are obtained for a fault-free scenario by employing the fault-tolerant routing technique mentioned in \cite{TCAD15}.  Even the other benchmark applications exhibit similar trends in the link utilization ratios. These statistics aid the adversary in locating the links to place the HTs in the NoC. For instance, the attacker may insert the HTs inside router 35 or router 36, as the number of packets utilizing the links associated with these routers exceeds 9\% of the total number of packets in the NoC. Thus, if an HT present in any of the router mentioned above triggers, it can drop 9\% of the total packets in the NoC, which can be disruptive in critical systems.

\begin{figure}[!ht]
\centering
\begin{tikzpicture}
\begin{axis}[
ybar,
bar width = 1pt,
xmin = -2,
xmax = 65,
width = 0.5\textwidth,
height = 0.325\textwidth,
xlabel = Router Index,
ylabel = Traffic Injection/Ejection (\%),
xtick pos = left,
]
    \addplot[fill=black!100] plot table {blackscholes_RUR_ForPlots.txt};
\end{axis}
\end{tikzpicture}
\caption{Traffic injection/ejection ratio of routers in an 8$\times$8 NoC when running \textit{blackscholes} benchmark.}
\label{fig:RUR}
\end{figure}
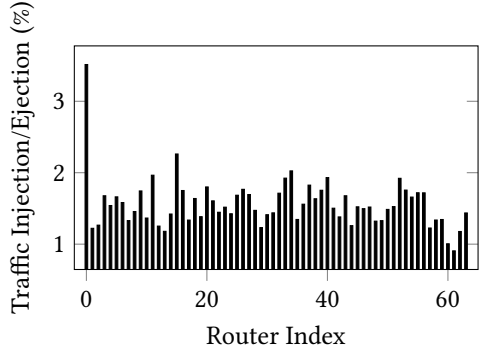

    The link utilization ratios shown in Fig. \ref{fig:linkutilblack} are for a fault-free case of the NoC. When a link in the NoC becomes faulty, these values deviate from the ones shown in the figure. Further, the routing algorithm considered in the current analysis adopts a look-ahead mechanism in deciding the paths for the packets. This look-ahead mechanism may avoid the routes that contain the routers associated with the faulty links, through which the packets had traveled when there were no link faults. Thus, to still attack with a high disruption rate, the adversary may consider the router utilization ratios to locate the points of insertions of the HTs. Fig. \ref{fig:RUR} shows the amount of injected/ejected traffic of each router in an 8$\times$8 NoC when running the \textit{blackscholes} application. As these traffic values remain constant for a mapped application on the NoC, an attack raised in a router would drop these packets, if not all packets that pass through the router. For instance, if a packet drop attack is successfully raised for the shown benchmark in router 35, the minimum and maximum packet loss will be 1.35\% and 9.24\% of the total number of packets, respectively.

\begin{figure}[!ht]
\centering
\begin{tikzpicture}
\begin{axis}[
ybar,
width = \textwidth,
height = 0.25\textwidth,
ymin = 0,
xtick = data,
xticklabel style={rotate=45},
x label style={at={(axis description cs:0.5,-1.0)},anchor=north},
area legend,
symbolic x coords = {blackscholes, canneal, dedup, ferret, fluidanimate, freqmine, raytrace, vips, x264},
enlarge x limits = 0.075,
legend style = {at = {(0.35,1.0)}, anchor = north, legend columns = 3},
ylabel = Packet Drop (\%),
xlabel = Benchmark Application,
xtick pos = left,
]
\addplot[draw=black!67, fill=black!67] coordinates {
(blackscholes,1.279295)
(canneal,1.562632)
(dedup,0.818525)
(ferret,0.803703)
(fluidanimate,1.224651)
(freqmine,0.286503)
(raytrace,1.462754)
(vips,0.841853)
(x264,0.408678)
};
\addlegendentry{Minimum}
\addplot[draw=black, fill=black] coordinates {
(blackscholes,6.146234)
(canneal,6.232109)
(dedup,5.148736)
(ferret,4.763734)
(fluidanimate,7.573920)
(freqmine,5.832338)
(raytrace,5.571316)
(vips,4.510910)
(x264,5.647979)
};
\addlegendentry{Average}
\addplot[draw=black!33, fill=black!33] coordinates {
(blackscholes,10.649730)
(canneal,9.566589)
(dedup,9.824388)
(ferret,7.911704)
(fluidanimate,16.938833)
(freqmine,34.465580)
(raytrace,8.504024)
(vips,7.855396)
(x264,24.310722)
};
\addlegendentry{Maximum}
\end{axis}
\end{tikzpicture}
\caption{Packet drop ratio (\%) of PARSEC benchmark applications when run on an 8$\times$8 NoC with SeFaR. Full-system simulation has been run for the NoC with a maximum of four HTs in the NoC that affects four out of five input ports of a router.}
\label{fig:PacketDropRatio}
\end{figure}
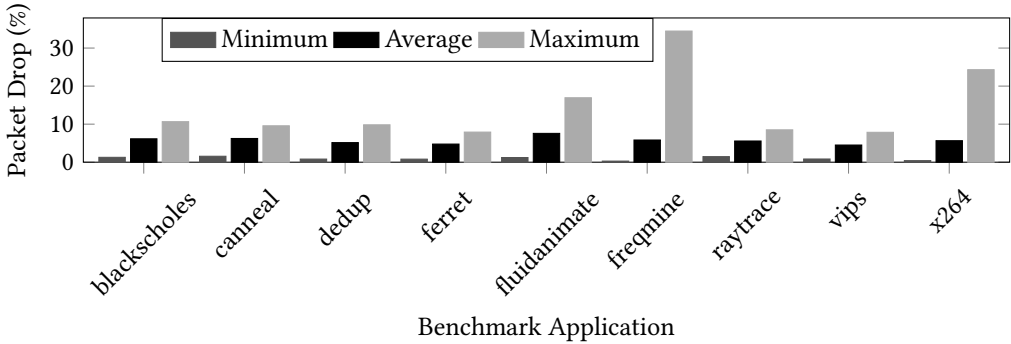

    Further, Fig. \ref{fig:PacketDropRatio} shows the packet drop ratio trends of various PARSEC benchmark applications for an 8$\times$8 mesh NoC when one of the routers is infected with the HTs. From the figure, one may note that the maximum packet drop ratio of an application can go as high as 34.47\%, with only one infected router in the NoC. These values, which show the severity of the attack, would further increase if multiple routers become victims of the packet drop attacks.

\subsection{Design Features of the HT}
\label{sub:threatmodel}
\begin{figure}[!ht]
\centering
\subfloat[]{\label{fig:lht}\includegraphics[width=0.33\textwidth]{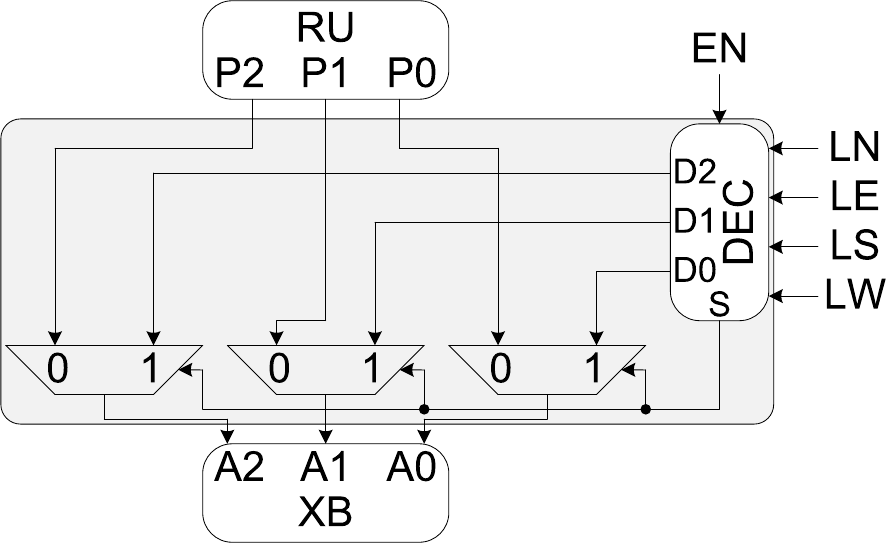}} \hspace{5pt}
\subfloat[]{\label{fig:lddec}\includegraphics[width=0.25\textwidth]{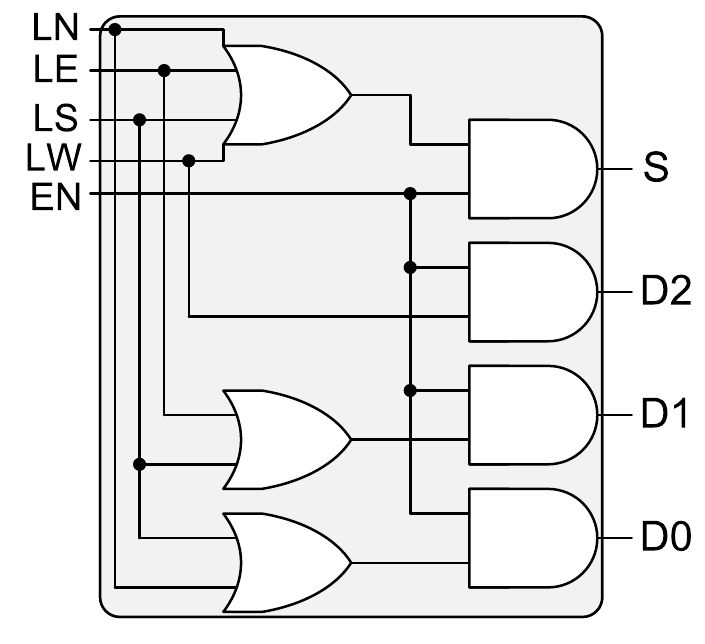}} \hspace{5pt}
\subfloat[]{\label{fig:ttdec}\resizebox{0.33\textwidth}{!}{
        \begin{tabular}[b]{|c|c|c|c|c|c|c|c|c|}
            \hline
            EN & LN & LE & LS & LW & D2 & D1 & D0 & S \\ \hline
            0 & $\times$ & $\times$ & $\times$ & $\times$ & 0 & 0 & 0 & 0 \\ \hline
            1 & 0 & 0 & 0 & 0 & 0 & 0 & 0 & 0 \\ \hline
            1 & 0 & 0 & 0 & 1 & 1 & 0 & 0 & 1 \\ \hline
            1 & 0 & 0 & 1 & 0 & 0 & 1 & 1 & 1 \\ \hline
            1 & 0 & 1 & 0 & 0 & 0 & 1 & 0 & 1 \\ \hline
            1 & 1 & 0 & 0 & 0 & 0 & 0 & 1 & 1 \\ \hline
        \end{tabular}}}
\caption{(a) Logic of considered hardware Trojan (HT). (b) Logic of DEC. (c) Truth table of DEC.}
\end{figure}

    For an HT to be effective, its area and power parameters ought to be insignificant, when compared to those corresponding to the conventional design in which they are placed. Fig. \ref{fig:lht} shows a design of an HT of the proposed attack. The HT consists of a link-to-port decoder (DEC) that decodes the port index corresponding to the faulty link, and three multiplexers to select the desired port index once the attack is created. For this HT, the DEC block serves as the trigger, and the three multiplexers serve as the payload. Inputs to the DEC are status signals of four directional links of a router associated with north (LN), east (LE), south (LS), and west (LW) ports, respectively, and an enable (EN) signal, which activates the DEC. The EN signal is a backdoor kill switch, similar to the one mentioned in \cite{Boraten2017}. The EN signal can also be realized using FSM based designs, to make the activation mechanism more complex to be detected. Other inputs to the HT constitute the input port index (P2P1P0) that serves as outputs to the RU. The outputs of the HT are the actual port index (A2A1A0) that serves as inputs to the crossbar and the select signal (S) to the multiplexers. The port indices of 001, 010, 011, and 100 correspond to the north (N), east (E), west (W), and south (S) ports, respectively. The purpose of the DEC block is to change the input port index to the compromised port index (D2D1D0) to trigger the attack, when EN = `1' and any of the link status is `1'. Figs. \ref{fig:lddec} and \ref{fig:ttdec} show the hardware of the DEC block in a proposed HT and its corresponding truth table.
    
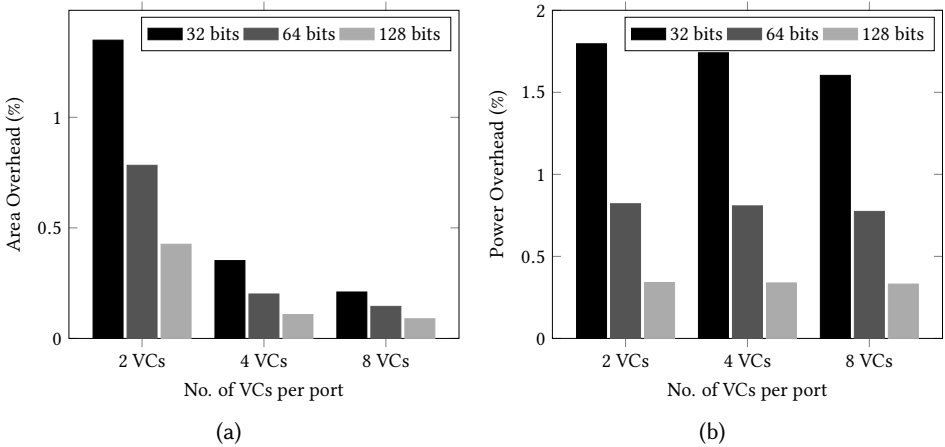
\begin{figure}[!ht]
\centering
\subfloat[]{\label{htav}
\resizebox{0.45\textwidth}{!}{
\begin{tikzpicture}
                \begin{axis}[ybar,
                    xtick = data,
                    ytick = {0, 0.5, 1, 1.5},
                    symbolic x coords = {2 VCs,4 VCs,8 VCs},
                    xlabel = No. of VCs per port,
                    ylabel = Area Overhead (\%),
                    bar width = 15pt,
                    enlarge x limits = 0.3,
                    ymin = 0,
                    area legend,
                    legend columns = 3,
                ]
                    \addplot [draw=black!100, fill=black!100] coordinates {
                        (2 VCs,1.349)
                        (4 VCs,0.352)
                        (8 VCs,0.210)
                    };
                    \addlegendentry{32 bits}
                    \addplot [draw=black!67, fill=black!67] coordinates {
                        (2 VCs,0.783)
                        (4 VCs,0.201)
                        (8 VCs,0.145)
                    };
                    \addlegendentry{64 bits}
                    \addplot [draw=black!33, fill=black!33] coordinates {
                        (2 VCs,0.426)
                        (4 VCs,0.108)
                        (8 VCs,0.089)
                    };
                    \addlegendentry{128 bits}
                \end{axis}
            \end{tikzpicture}
}
            }
\subfloat[]{\label{htpv}
\resizebox{0.45\textwidth}{!}{
\begin{tikzpicture}
                \begin{axis}[ybar,
                    xtick = data,
                    symbolic x coords = {2 VCs,4 VCs,8 VCs},
                    xlabel = No. of VCs per port,
                    bar width = 15pt,
                    enlarge x limits = 0.3,
                    ylabel = Power Overhead (\%),
                    ymin = 0, ymax = 2,
                    area legend,
                    legend columns = 3,
                ]
                    \addplot [draw=black!100, fill=black!100] coordinates {
                        (2 VCs,1.796)
                        (4 VCs,1.741274244)
                        (8 VCs,1.603380214)
                    };
                    \addlegendentry{32 bits}
                    \addplot [draw=black!67, fill=black!67] coordinates {
                        (2 VCs,0.8209690921)
                        (4 VCs,0.8080223798)
                        (8 VCs,0.7742699485)
                    };
                    \addlegendentry{64 bits}
                    \addplot [draw=black!33, fill=black!33] coordinates {
                        (2 VCs,0.3410846607)
                        (4 VCs,0.3383173975)
                        (8 VCs,0.3312717729)
                    };
                    \addlegendentry{128 bits}
                \end{axis}
            \end{tikzpicture}
}
            }
\caption{(a) Area overhead and (b) Power overhead of the proposed HT with respect to baseline fault-tolerant NoC router.}
\end{figure}

    Fig. \ref{htav} and \ref{htpv} show the design foot-print of proposed HT regarding both percentage area overhead and percentage power overhead when compared to the baseline FT router. The values have been obtained by synthesizing the modules using Synopsys DC and Faraday 90 nm library, for an operating voltage of 1 V and an operating frequency of 1 GHz. Area and power overheads are calculated by considering five HTs in a router. Each router has five IO ports, wherein each virtual channel (VC) is capable of storing four flits. Size of the NoC considered is 8$\times$8, and its topology is mesh. From the figures, it is evident that both area and power overheads of HTs are less than 1.35\% and 1.79\%, respectively, for a router with 2 VCs per IO port and flit width of 32 bits. As routers constitute only a small fraction of the area in a NoC based MPSoC, 1.35\% area overhead and 1.79\% power overhead are insignificantly low values when compared with those of the entire NoC based MPSoC. Further, the area and power overheads reduce with an increase in the flit width of the NoC. This comparison with different flit widths is to show that the design footprint of the HTs does not depend on the flit width of the network. Another observation one can make from the plots is as the share of the design footprint of the HTs reduces the complexity of their detection in such designs increases.
    
\section{Proposed Mitigation Mechanism}
\label{sec:proposed}
\begin{figure}[!ht]
    \centering
        \includegraphics[width=0.35\textwidth]{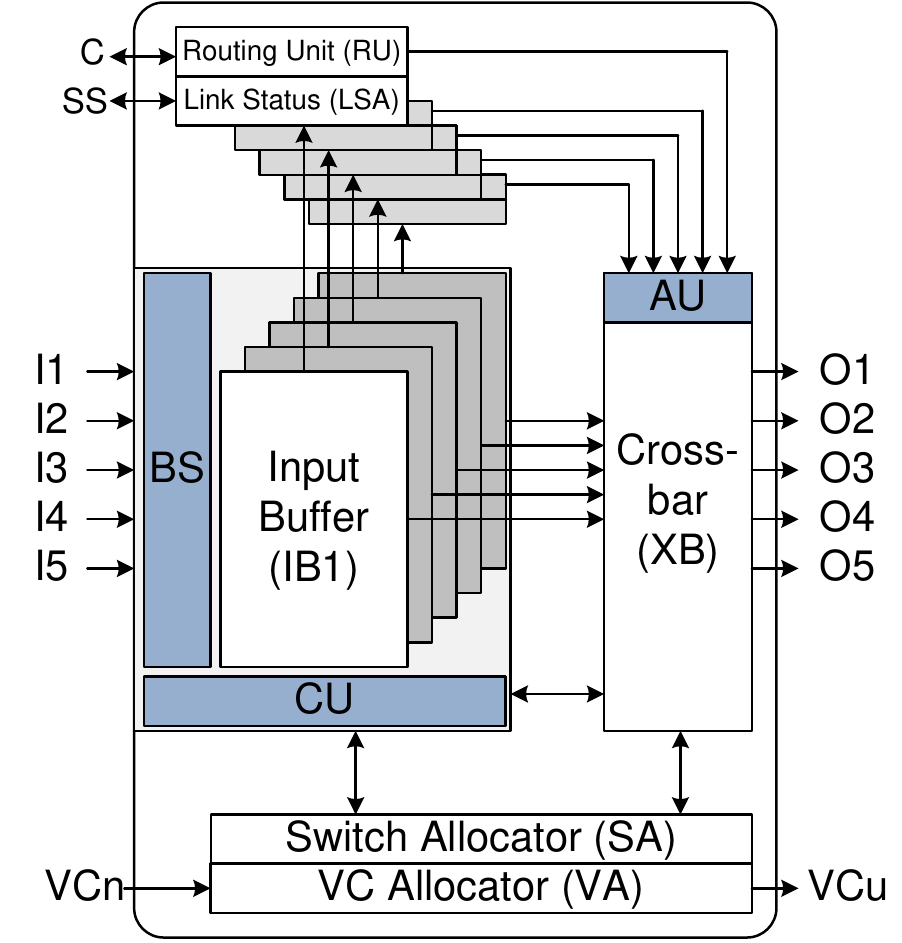}
        \caption{Microarchitecture of the proposed secure fault-tolerant router (SeFaR).}
        \label{fig:sefar}
    \end{figure}

    To counter the consequences caused because of the proposed HTs, a secure FT router, called SeFaR, has been proposed. To establish security within SeFaR, security components, namely the authentication unit (AU), the control unit (CU), and the buffer shuffler (BS) have been introduced. Fig. \ref{fig:sefar} shows the microarchitecture of SeFaR along with the blocks mentioned above. The AU block connects with the XB and acts as a shield by blocking the port indices, which attempt to force the packets injected into the router to proceed to the output ports containing faulty links. CU controls the BS block in securing the router, by considering the output signals of the AU. BS shuffles the buffer locations where the input data needs to occupy.
    
\subsection{Buffer Shuffler (BS)}
\label{sub:BS}
    The purpose of the BS block, in SeFaR, is to shuffle the port indices such that the incoming data is sent into a buffer, whose routing unit is not compromised. Design of the BS block is straightforward, as it conforms to the logic of a crossbar. However, to keep its logic simple, pass transistor based crossbar can be considered instead of a multiplexer based one. A router with $N$ ports requires an $N \times N$ BS. The select inputs to the BS come from the CUs of respective ports in the router.
    
\subsection{Authentication Unit (AU)}
\label{sub:AU}
\begin{figure}[!ht]
\centering
\subfloat[]{\label{fig:ttau}\resizebox{0.33\textwidth}{!}{
        \begin{tabular}[b]{|c|c|c|c|c|c|c|c|}
                \hline
                LN & LE & LS & LW & A2 & A1 & A0 & W \\ \hline
                0 & 0 & 0 & 0 & $\times$ & $\times$ & $\times$ & 0 \\ \hline
                0 & 0 & 0 & 1 & 1 & 0 & 0 & 1 \\ \hline
                0 & 0 & 1 & 0 & 0 & 1 & 1 & 1 \\ \hline
                0 & 1 & 0 & 0 & 0 & 1 & 0 & 1 \\ \hline
                1 & 0 & 0 & 0 & 0 & 0 & 1 & 1 \\ \hline
        \end{tabular}}}
\subfloat[]{\label{fig:ldau}\includegraphics[width=0.21\textwidth]{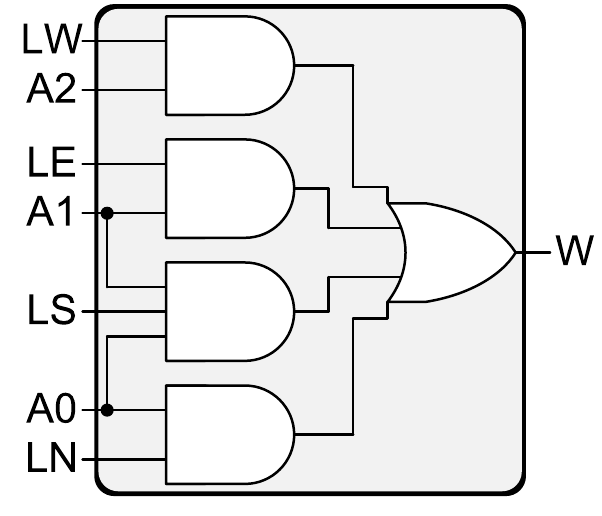}}
\subfloat[]{\label{fig:lcu}\includegraphics[width=0.45\textwidth]{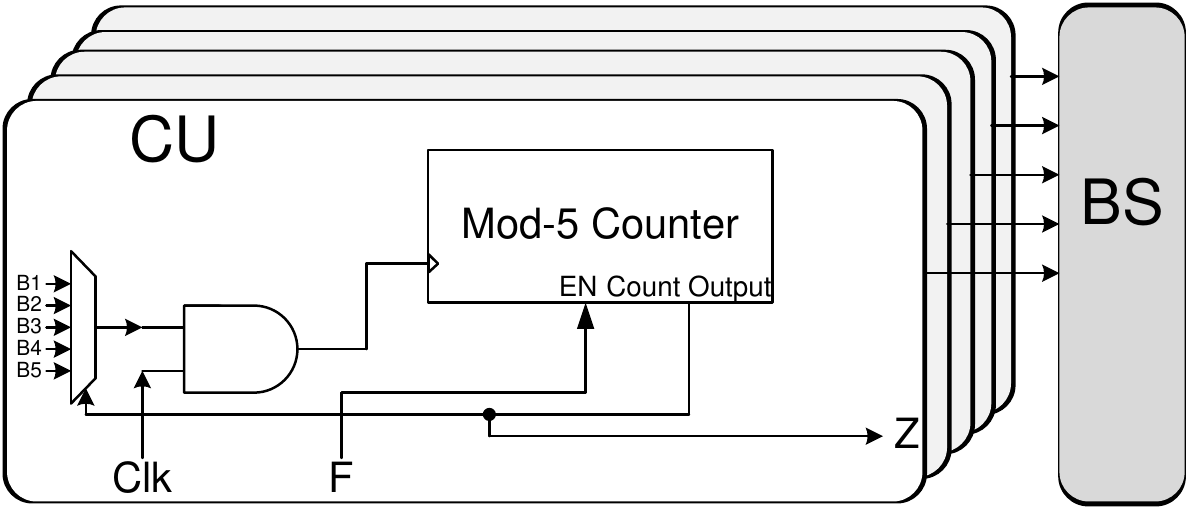}}
\caption{(a) Truth table of the proposed authentication unit (AU). (b) Logic of the proposed authentication unit (AU). (c) Logic of the proposed Control Unit (CU).}
\end{figure}

    Workflow of the logic corresponding to the AU has been summarized next. AU continuously gathers the link status information from the LSA and keeps on checking for any anomaly in the port indices obtained from the routing units. If any anomaly is detected, it raises a warning flag (F) associated with a particular port to `1'. This flag indicates the presence of an HT in the router at the routing unit (RU) associated with that port and alerts the CU to take further action. Figs. \ref{fig:ttau} and \ref{fig:ldau} show the truth table and logic diagram of the proposed AU.
    
\subsection{Control Unit (CU)}
\label{sub:CU}
    Fig. \ref{fig:lcu} shows the logic of CU and its association with the BS block. If there are $N$ ports in a router, there would be $N$ CU blocks, each corresponding to a port of the router. Since CU generates the state bits to be given to the BS block, an FSM-based modulo-N counter has been employed, with the initial states being the port indices. Here, B1 to B5 represent the busy status flags of the buffers of all ports. A logic `1' of any of these means that the buffer has either been occupied by other CU. This condition indicates that no two CUs should have same states in their FSMs, else the RU associated with the corresponding port has been compromised. $F$ is the warning signal coming from AU, and $Z$ is the output of the FSM, which serves as the input to the BS block.
    
    The functionality of the CU can be explained as follows. Initially, the seed value corresponding to the port is set, and the FSM outputs the same, which directs the shuffler to forward the incoming packets to their default buffers. The same output works as the select signal for the multiplexer present in CU, which selects the buffer status signal corresponding to its port. It then waits for $F$ of respective port, to enable the FSM to change its output, thus asserting the shuffler to redirect the incoming packets to some other buffer locations. The signal $F$ acts as the enable signal to CU. Once $F$ of a port becomes high, the buffer status flag also becomes high, which makes the FSM change its state, transiting to its next state. Then, depending on the availability of the buffers of the port corresponding to its new state, the FSM decides whether to change its state further or to remain in that state.
    
\begin{figure}[!ht]
        \centering
        \includegraphics[width=0.4825\textwidth]{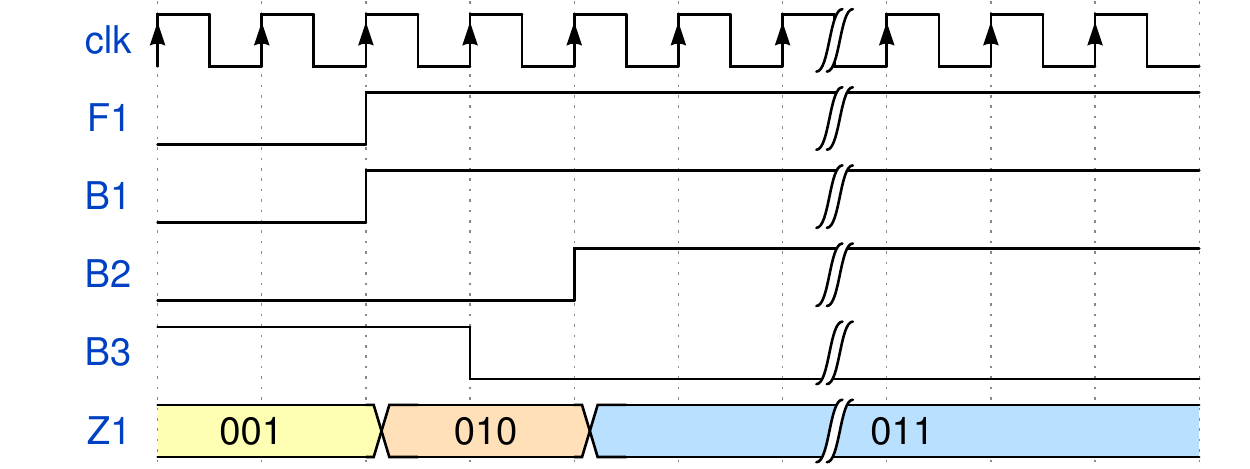}
        \caption{Timing diagram showing functionality of the CU.}
        \label{fig:timingCU}
    \end{figure}

    An example to illustrate the functionality of the CU is shown in Fig. \ref{fig:timingCU}. Here, port I1 is considered, which is associated with the signals $B1$ and $F1$. The initial seed, as well as the output of the constituent FSM, is 001. From the diagram, as soon as $F1$ goes high, B1 follows it, cautioning the CU to change its state. CU changes its state from 001 to 010, which corresponds to P2. Now, the FSM checks for the availability of associated buffer, by checking the B2 flag. $Z1$ remains in state 010, till $B2$ remains zero. This condition means that the shuffler now allows the packets coming from I1 to occupy the input buffer IB2. As soon as the flag $B2$ goes high, $Z1$ changes to its next free state. This process continues until BS finds a suitable port for the packets to occupy.
    
    The proposed CU can also be designed using look-up tables (LUTs), instead of FSMs, if the application traffic is known beforehand. Also, with the emergence of dynamic VC management techniques, as in \cite{dynamicvc}, the proposed CU can be integrated with the VC management unit of such techniques, to enhance the serviceability of the NoC. However, the difference between the dynamic VC management and the functionality of the CU is that the dynamic VC management manages the allocation of the VCs present inside a buffer at an input port, whereas the CU manages the allocation of the buffers itself.

\subsection{An Example Scenario}
\label{sub:example}
\begin{figure}[!ht]
\centering
\includegraphics[width = \textwidth]{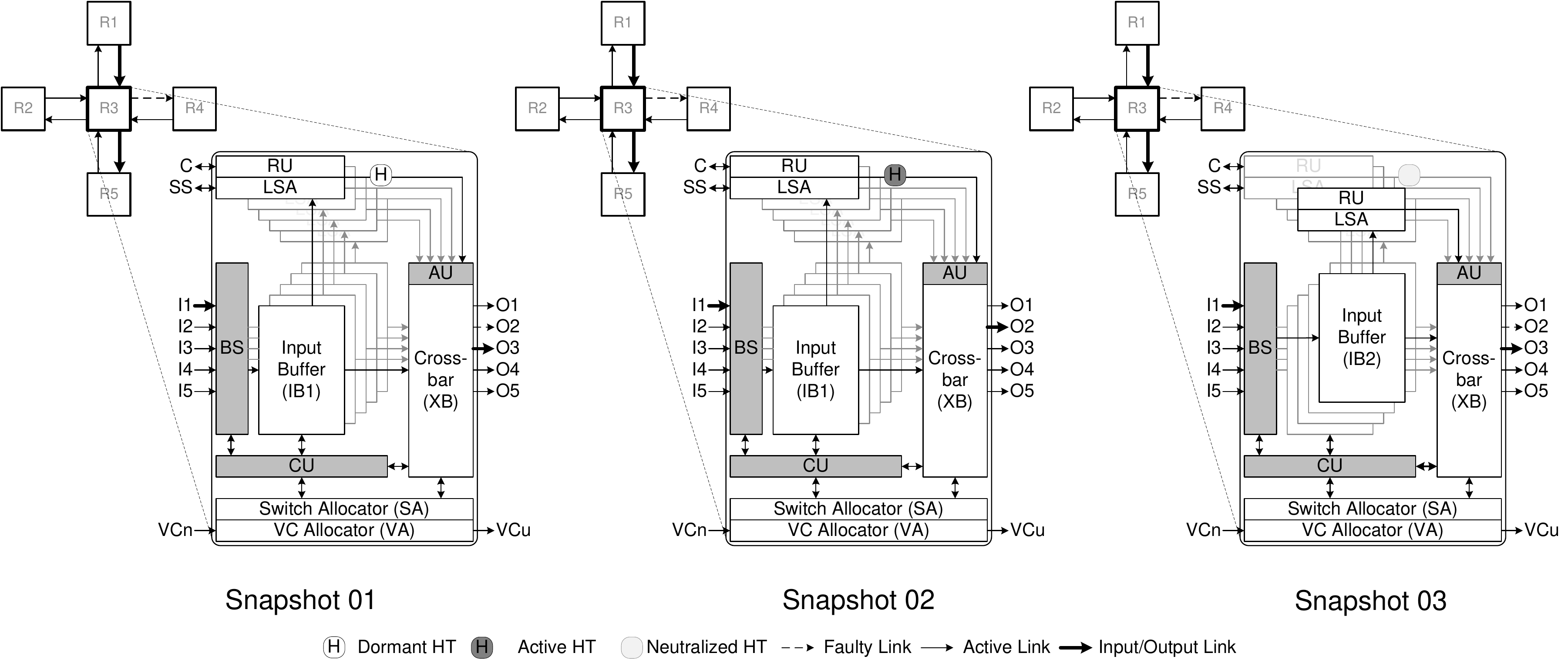}
\caption{Snapshots of a scenario showing the execution steps of SeFaR in the event of a packet drop attack.}
\label{fig:example}
\end{figure}

    A better understanding of how the proposed runtime mitigation technique using SeFaR functions can be obtained using an example. Fig. \ref{fig:example} shows the snapshots of a scenario, wherein the execution steps of SeFaR are shown. The following steps elaborate the execution of SeFaR in the event of a packet drop attack. The scenario is that a packet has to travel from the router R1 to the router R5 via the router R3. The input and output ports through which the packet passes through R3 are I1 and O3, respectively. The outbound link associated with the port connecting R3 and R4, namely O2, has become faulty and there exists an HT that can raise a packet drop attack in R3, near the routing unit of the input port I1. The default data path of the packet through the central router is the input port I1, the buffer shuffler (BS), the input buffer IB1, the crossbar, and the output port O3. It is assumed that there is no contention in R3 for the packets arriving at I1 due to the other ports.

    \begin{itemize}
        \item {\textbf{Snapshot 01} At this instance, the HT present at the routing unit of the input port I1 is not triggered. Thus the packet follows the default path and occupies the input buffer B1. With the knowledge of the faulty link associated with the output port O2, the routing unit of input port I1 does not forward any packet through this output port.}
        \item {\textbf{Snapshot 02} At this instance, the HT present at the routing unit of the input port I1 triggers, and manipulates the output port indices obtained from the routing unit, so that the packet is forced to go through the output port O2, instead of O3. This situation makes the router R3 a packet drop router, with respect to the input port I1. AU, present at the crossbar, finds that the packet from the input port I1 is being forced to the output port O2, and raises a warning flag and communicates with the CU, located at the input buffer IB1.}
        \item {\textbf{Snapshot 03} At this instance, the CU associated with the input buffers applies the permutation and shuffles the buffer index such that the packets arriving at the input port I1, which had to occupy B1, now occupy B2. Simultaneously, the CU instructs the switch allocator not to consider further packet requests from the input buffer IB1. Since the RU associated with the input buffer IB2 is not affected by the existing Trojan, it forwards the packets to their legitimate output ports, which is O3 in this case.}
    \end{itemize}
    
    SeFaR thus restores the communication flow by preventing the network from losing the packets at the router R3. The new path for the packets from the router R1 to the router R5 through the router R3 will be the input port I1, the input buffer IB2, the crossbar, and the output port O3.

\section{Performance Evaluation}
\label{sec:results}
Performance metrics such as area, power, and performance overheads of SeFaR have been compared with the baseline FT router. Architectures of both baseline and secure FT routers have been coded in VHDL. Synopsys Design Compiler \cite{Synopsys} and Faraday 90 nm technology library have been used in implementing the designs. An operating frequency of 1 GHz has been set for the routers. The size of the network considered is 8$\times$8, and the topology is mesh. Each router has five input/output (IO) ports, and each VC in a buffer can store four flits.

\subsection{Area and Power Overhead Evaluation}
\label{sub:areaov}
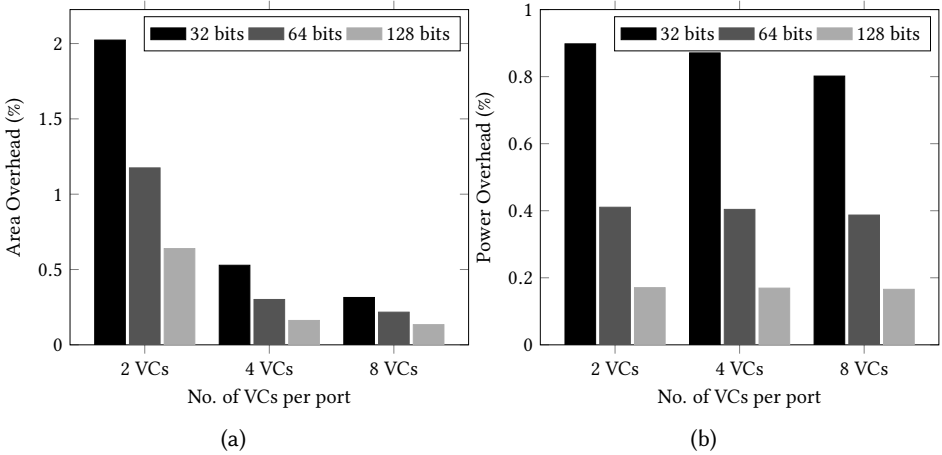
\begin{figure}[!ht]
\centering
\subfloat[]{\label{fig:audav}\resizebox{0.45\textwidth}{!}{\begin{tikzpicture}
\begin{axis}[ybar,
    xtick = data,
    symbolic x coords = {2 VCs,4 VCs,8 VCs},
    xlabel = No. of VCs per port,
    ylabel = Area Overhead (\%),
    ymin = 0,
    bar width = 15pt,
    enlarge x limits = 0.3,
    area legend,
    legend columns = 3,
]
    \addplot [draw=black!100, fill=black!100] coordinates {
        (2 VCs,2.023143496)
        (4 VCs,0.5283541861)
        (8 VCs,0.3144064454)
    };
    \addlegendentry{32 bits}
    \addplot [draw=black!67, fill=black!67] coordinates {
        (2 VCs,1.174645345)
        (4 VCs,0.3011303228)
        (8 VCs,0.2169778603)
    };
    \addlegendentry{64 bits}
    \addplot [draw=black!33, fill=black!33] coordinates {
        (2 VCs,0.6388136522)
        (4 VCs,0.1618871698)
        (8 VCs,0.1339569385)
    };
    \addlegendentry{128 bits}
\end{axis}
            \end{tikzpicture}}}
\subfloat[]{\label{fig:audpv}\resizebox{0.45\textwidth}{!}{\begin{tikzpicture}
\begin{axis}[ybar,
    xtick = data,
    symbolic x coords = {2 VCs,4 VCs,8 VCs},
    xlabel = No. of VCs per port,
    ylabel = Power Overhead (\%),
    ymin = 0, ymax = 1,
    bar width = 15pt,
    enlarge x limits = 0.3,
    area legend,
    legend columns = 3,
]
    \addplot [draw=black!100, fill=black!100] coordinates {
        (2 VCs,0.8978182514)
        (4 VCs,0.8707872121)
        (8 VCs,0.8016901071)
    };
    \addlegendentry{32 bits}
    \addplot [draw=black!67, fill=black!67] coordinates {
        (2 VCs,0.410484546)
        (4 VCs,0.4040111899)
        (8 VCs,0.3871349743)
    };
    \addlegendentry{64 bits}
    \addplot [draw=black!33, fill=black!33] coordinates {
        (2 VCs,0.1705423304)
        (4 VCs,0.1691586988)
        (8 VCs,0.1656358865)
    };
    \addlegendentry{128 bits}
\end{axis}
            \end{tikzpicture}}}
\caption{(a) Area overhead and (b) Power overhead of secure fault-tolerant router (SeFaR) compared to baseline fault-tolerant router.}
\end{figure}

Fig. \ref{fig:audav} shows the area overhead (in \%) of SeFaR when compared with the baseline FT router. To evaluate the metric over a wide range, we considered routers with two, four, and eight virtual channels (VCs) per IO port. For each of these configurations, the flit width has been varied from 32 bits to 128 bits. We have synthesized the SeFaR with five AUs in each router. From the figure, it is evident that for a 32-bit flit in a router with 2 VCs per IO port, the area overhead of SeFaR is found to be 2.02\%. This overhead is further reduced when higher configurations of routers are considered.
    
Fig. \ref{fig:audpv} shows the power overhead (in \%) of SeFaR when compared with the baseline FT router. Since the additional logic associated with SeFaR is very little, the power overhead achieved by it is also very less. From the figure, for SeFaR with 32-bit wide flits and 2 VCs per IO port, the power overhead is 0.90 \%. This overhead is further reduced when higher configurations of routers are considered. When this power overhead is measured with respect to the power consumption of NoC based MPSoC, it becomes insignificantly low, which makes SeFaR a promising alternative to baseline FT routers, with an additional security layer.
    
\subsection{Performance Overhead Evaluation}
\label{sub:performance}
To evaluate the performance overhead of SeFaR and compare it with that of the baseline FT router, we have obtained the traffic distributions of applications from the PARSEC benchmark suite using the SniperSim \cite{sniper} full-system simulator. Further, to analyze the performance of SeFaR for synthetic traffic, patterns such as shuffle, transpose, and uniform-random (uniform) have been considered. The network evaluation for all real, as well as synthetic, traffic distributions, has been done using the BookSim \cite{ispass} network simulator. Performance overhead evaluation has been done for the metrics such as execution time and energy consumption overheads for real benchmark applications. For synthetic traffic, parameters such as average packet latency (APL) and power-latency product (PLP) have been analyzed for SeFaR and compared with that of the baseline FT router. An advantage of analyzing the performance of SeFaR for synthetic traffic is that one can estimate the overheads of the mitigation mechanism for a range of packet injection rates (PIRs).

\subsubsection{Real Benchmarks}
\label{subsub:real}
For the current evaluation with real benchmark traffic patterns, we considered one packet drop router in the network at a time. Further, to observe the system performance for different routers acting as packet drop routers, results have been obtained for all the cases one after the other. A packet drop router in the current scenario is analyzed for two cases---one with a single HT that affects one input port, the other with four HTs that affect four input ports.

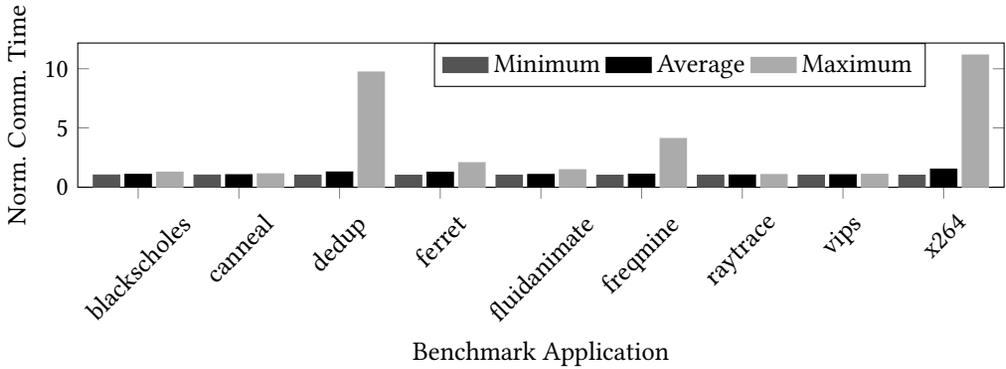
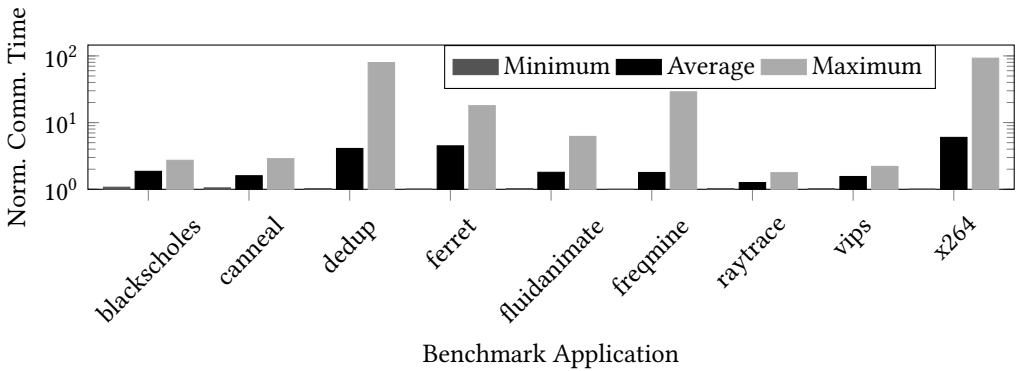
\begin{figure}[!ht]
\centering
\subfloat[]{\label{fig:CTOV1HT}
\begin{tikzpicture}
\begin{axis}[
ybar,
width = \textwidth,
height = 0.25\textwidth,
xtick = data,
xticklabel style={rotate=45},
x label style={at={(axis description cs:0.5,-1.0)},anchor=north},
area legend,
symbolic x coords = {blackscholes, canneal, dedup, ferret, fluidanimate, freqmine, raytrace, vips, x264},
enlarge x limits = 0.075,
legend style = {at = {(0.65,1.0)}, anchor = north, legend columns = 3},
ylabel = Norm. Comm. Time,
xlabel = Benchmark Application,
xtick pos = left,
]
\addplot[draw=black!67, fill=black!67] coordinates {
(blackscholes,1.018743)
(canneal,1.012747)
(dedup,1.006107)
(ferret,1.000823)
(fluidanimate,1.005194)
(freqmine,1.000000)
(raytrace,1.006122)
(vips,1.004086)
(x264,1.000388)
};
\addlegendentry{Minimum}
\addplot[draw=black, fill=black] coordinates {
(blackscholes,1.069082)
(canneal,1.043169)
(dedup,1.274189)
(ferret,1.250584)
(fluidanimate,1.063149)
(freqmine,1.086824)
(raytrace,1.020100)
(vips,1.040029)
(x264,1.516532)
};
\addlegendentry{Average}
\addplot[draw=black!33, fill=black!33] coordinates {
(blackscholes,1.257427)
(canneal,1.117342)
(dedup,9.725488)
(ferret,2.058807)
(fluidanimate,1.465169)
(freqmine,4.103117)
(raytrace,1.053239)
(vips,1.075246)
(x264,11.161947)
};
\addlegendentry{Maximum}
\end{axis}
\end{tikzpicture}
}\\
\subfloat[]{\label{fig:CTOV4HT}
\begin{tikzpicture}
\begin{axis}[
ybar,
ymode = log,
ymin = 1,
width = \textwidth,
height = 0.25\textwidth,
xtick = data,
xticklabel style={rotate=45},
x label style={at={(axis description cs:0.5,-1.0)},anchor=north},
area legend,
symbolic x coords = {blackscholes, canneal, dedup, ferret, fluidanimate, freqmine, raytrace, vips, x264},
enlarge x limits = 0.075,
legend style = {at = {(0.65,1.0)}, anchor = north, legend columns = 3},
ylabel = Norm. Comm. Time,
xlabel = Benchmark Application,
xtick pos = left,
]
\addplot[draw=black!67, fill=black!67] coordinates {
(blackscholes,1.074972)
(canneal,1.050990)
(dedup,1.024427)
(ferret,1.003292)
(fluidanimate,1.020777)
(freqmine,1.000000)
(raytrace,1.027578)
(vips,1.016345)
(x264,1.001552)
};
\addlegendentry{Minimum}
\addplot[draw=black, fill=black] coordinates {
(blackscholes,1.860119)
(canneal,1.597664)
(dedup,4.096079)
(ferret,4.481721)
(fluidanimate,1.798682)
(freqmine,1.787540)
(raytrace,1.262343)
(vips,1.555532)
(x264,6.002273)
};
\addlegendentry{Average}
\addplot[draw=black!33, fill=black!33] coordinates {
(blackscholes,2.721061)
(canneal,2.877469)
(dedup,79.529404)
(ferret,17.940912)
(fluidanimate,6.201559)
(freqmine,28.928061)
(raytrace,1.775564)
(vips,2.203932)
(x264,92.457542)
};
\addlegendentry{Maximum}
\end{axis}
\end{tikzpicture}
}
\caption{Communication time overhead of PARSEC benchmark applications when run on an 8$\times$8 NoC with SeFaR. Full-system simulation has been run for the NoC with (a) one infected router in the NoC with one HT and (b) one infected router in the NoC with a maximum of four HTs in the NoC that affects four out of its five input ports.}
\end{figure}

Figs. \ref{fig:CTOV1HT} and \ref{fig:CTOV4HT} show a comparison of SeFaR with the baseline FT router regarding the overheads observed in the communication time while running the real benchmarks. Fig. \ref{fig:CTOV1HT} refers to the NoC with one router that is infected with a single HT that drops the packets arriving at an input port, whereas Fig. \ref{fig:CTOV4HT} refers to the NoC with one router that is infected with a maximum of four HTs, which drop the packets arriving at all but one input ports of the router. From the plots, one can find that for the case of a router infected with a single HT, the overhead observed in the communication time can go as high as 10$\times$ the baseline value. Similarly, for the case of a router infected with a maximum of four HTs, the overheads observed in the communication time can go as high as 90$\times$ the baseline value. Despite the huge overheads accounted in SeFaR to mitigate the packet drop attacks while running the PARSEC benchmarks, the application execution time has not been observed to increase similarly. This increasing trend in the application execution time is because of the core idle time and the dependence of the spatial and temporal traffic distributions of the application under execution. Details on the same are presented next. 

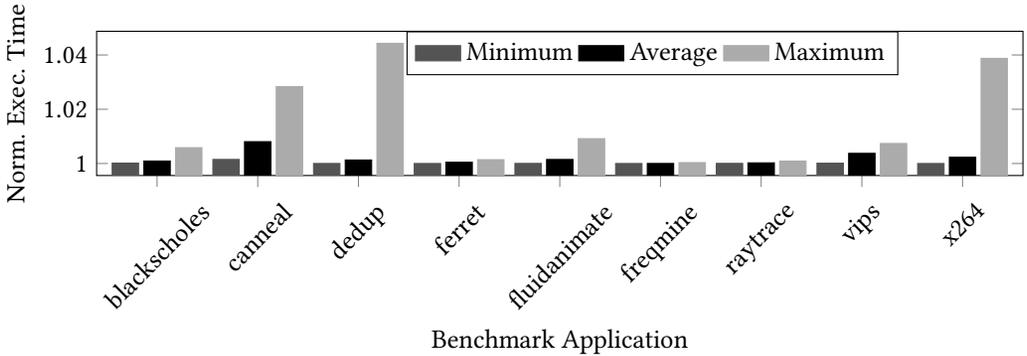
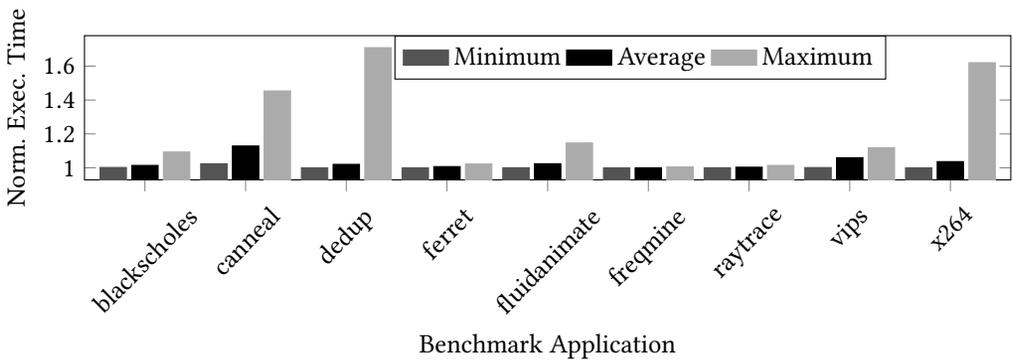
\begin{figure}[!ht]
\centering
\subfloat[]{\label{fig:ETO1Port}
\begin{tikzpicture}
\begin{axis}[
ybar,
width = \textwidth,
height = 0.25\textwidth,
xtick = data,
xticklabel style={rotate=45},
x label style={at={(axis description cs:0.5,-1.0)},anchor=north},
area legend,
symbolic x coords = {blackscholes, canneal, dedup, ferret, fluidanimate, freqmine, raytrace, vips, x264},
enlarge x limits = 0.075,
legend style = {at = {(0.6,1.0)}, anchor = north, legend columns = 3},
ylabel = Norm. Exec. Time,
xlabel = Benchmark Application,
xtick pos = left,
]
    \addplot [draw=black!67, fill=black!67] coordinates {
        (blackscholes,1.000156)
        (canneal,1.001499)
        (dedup,1.000007)
        (ferret,1.000005)
        (fluidanimate,1.000033)
        (freqmine,1.000000)
        (raytrace,1.000046)
        (vips,1.000120)
        (x264,1.000001)
    };
    \addlegendentry{Minimum}
    \addplot [draw, fill] coordinates {
        (blackscholes,1.000911)
        (canneal,1.008057)
        (dedup,1.001295)
        (ferret,1.000473)
        (fluidanimate,1.001493)
        (freqmine,1.000013)
        (raytrace,1.000243)
        (vips,1.003742)
        (x264,1.002311)
    };
    \addlegendentry{Average}
    \addplot [draw=black!33, fill=black!33] coordinates {
        (blackscholes,1.005846)
        (canneal,1.028360)
        (dedup,1.044299)
        (ferret,1.001419)
        (fluidanimate,1.009163)
        (freqmine,1.000334)
        (raytrace,1.000900)
        (vips,1.007394)
        (x264,1.038753)
    };
    \addlegendentry{Maximum}
\end{axis}
\end{tikzpicture}
}\\
\subfloat[]{\label{fig:ETO4Port}
\begin{tikzpicture}
\begin{axis}[
ybar,
width = \textwidth,
height = 0.25\textwidth,
xtick = data,
xticklabel style={rotate=45},
x label style={at={(axis description cs:0.5,-1.0)},anchor=north},
area legend,
symbolic x coords = {blackscholes, canneal, dedup, ferret, fluidanimate, freqmine, raytrace, vips, x264},
enlarge x limits = 0.075,
legend style = {at = {(0.6,1.0)}, anchor = north, legend columns = 3},
ylabel = Norm. Exec. Time,
xlabel = Benchmark Application,
xtick pos = left,
]
    \addplot [draw=black!67, fill=black!67] coordinates {
        (blackscholes,1.002503)
        (canneal,1.023981)
        (dedup,1.000116)
        (ferret,1.000073)
        (fluidanimate,1.000534)
        (freqmine,1.000000)
        (raytrace,1.000743)
        (vips,1.001923)
        (x264,1.000024)
    };
    \addlegendentry{Minimum}
    \addplot [draw, fill] coordinates {
        (blackscholes,1.014581)
        (canneal,1.128911)
        (dedup,1.020726)
        (ferret,1.007573)
        (fluidanimate,1.023884)
        (freqmine,1.000216)
        (raytrace,1.003891)
        (vips,1.059868)
        (x264,1.036967)
    };
    \addlegendentry{Average}
    \addplot [draw=black!33, fill=black!33] coordinates {
        (blackscholes,1.093537)
        (canneal,1.453762)
        (dedup,1.708793)
        (ferret,1.022709)
        (fluidanimate,1.146610)
        (freqmine,1.005348)
        (raytrace,1.014400)
        (vips,1.118300)
        (x264,1.620043)
    };
    \addlegendentry{Maximum}
\end{axis}
\end{tikzpicture}
}
\caption{Normalized execution time of PARSEC benchmark applications when run on an 8$\times$8 NoC with SeFaR. Full-system simulation has been run for the NoC with (a) one infected router in the NoC with one HT and (b) one infected router in the NoC with a maximum of four HTs in the NoC that affects four out of its five input ports.}
\end{figure}

Figs. \ref{fig:ETO1Port} and \ref{fig:ETO4Port} show a comparison of SeFaR with the baseline FT router regarding the overheads observed in the execution time while running the real benchmarks. Fig. \ref{fig:ETO1Port} refers to the routers that are infected with a single HT, whereas Fig. \ref{fig:ETO4Port} refers to the routers that are affected with four HTs. The links, which are faulty, have been considered to be faulty for the entire application execution. For the case of a single HT in a packet drop router, the maximum overhead in the execution time has been observed for \textit{dedup}, which is 4.46\%, when compared with the baseline FT router. However, the average overhead in the execution time for each application does not exceed 0.8\% with respect to the baseline FT router, when a packet drop router is infected with a single HT. Similarly, for the case with four HTs in a packet drop router, the maximum overhead observed for SeFaR regarding the execution time has been 70.87\% for \textit{dedup}, when compared with the baseline FT router. Whereas, the average overhead in the execution time for this case does not exceed 12.89\%, for the considered real benchmarks. The reason for the difference seen in the execution time overheads of these applications is because of the difference in their traffic distributions as well as traffic volumes.

\begin{figure}[!ht]
\centering
\subfloat[]{\label{fig:ENO1Port}
\begin{tikzpicture}
\begin{axis}[
ybar,
width = \textwidth,
height = 0.25\textwidth,
xtick = data,
xticklabel style={rotate=45},
x label style={at={(axis description cs:0.5,-1.0)},anchor=north},
area legend,
symbolic x coords = {blackscholes, canneal, dedup, ferret, fluidanimate, freqmine, raytrace, vips, x264},
enlarge x limits = 0.075,
legend style = {at = {(0.5,1.0)}, anchor = north, legend columns = 3},
ylabel = Normalized Energy,
xlabel = Benchmark Application,
xtick pos = left,
]
    \addplot [draw=black!67, fill=black!67] coordinates {
        (blackscholes,1.000000)
        (canneal,1.000000)
        (dedup,1.000000)
        (ferret,1.000000)
        (fluidanimate,1.000000)
        (freqmine,1.000000)
        (raytrace,1.000000)
        (vips,1.000000)
        (x264,1.000000)
    };
    \addlegendentry{Minimum}
    \addplot [draw, fill] coordinates {
        (blackscholes,1.000082)
        (canneal,1.283628)
        (dedup,1.001393)
        (ferret,1.000404)
        (fluidanimate,1.005712)
        (freqmine,1.000550)
        (raytrace,1.002616)
        (vips,1.207055)
        (x264,1.000306)
    };
    \addlegendentry{Average}
    \addplot [draw=black!33, fill=black!33] coordinates {
        (blackscholes,1.000226)
        (canneal,1.444494)
        (dedup,1.004128)
        (ferret,1.001253)
        (fluidanimate,1.011328)
        (freqmine,1.008806)
        (raytrace,1.004664)
        (vips,1.393188)
        (x264,1.003987)
    };
    \addlegendentry{Maximum}
\end{axis}
\end{tikzpicture}
}\\
\subfloat[]{\label{fig:ENO4Port}
\begin{tikzpicture}
\begin{axis}[
ybar,
width = \textwidth,
height = 0.25\textwidth,
xtick = data,
xticklabel style={rotate=45},
x label style={at={(axis description cs:0.5,-1.0)},anchor=north},
area legend,
symbolic x coords = {blackscholes, canneal, dedup, ferret, fluidanimate, freqmine, raytrace, vips, x264},
enlarge x limits = 0.075,
legend style = {at = {(0.5,1.0)}, anchor = north, legend columns = 3},
ylabel = Normalized Energy,
xlabel = Benchmark Application,
xtick pos = left,
]
    \addplot [draw=black!67, fill=black!67] coordinates {
        (blackscholes,1.000000)
        (canneal,1.000000)
        (dedup,1.000000)
        (ferret,1.000000)
        (fluidanimate,1.000000)
        (freqmine,1.000000)
        (raytrace,1.000000)
        (vips,1.000000)
        (x264,1.000000)
    };
    \addlegendentry{Minimum}
    \addplot [draw, fill] coordinates {
        (blackscholes,1.000088)
        (canneal,1.285880)
        (dedup,1.001466)
        (ferret,1.000380)
        (fluidanimate,1.005802)
        (freqmine,1.001051)
        (raytrace,1.002464)
        (vips,1.223087)
        (x264,1.000341)
    };
    \addlegendentry{Average}
    \addplot [draw=black!33, fill=black!33] coordinates {
        (blackscholes,1.000238)
        (canneal,1.821062)
        (dedup,1.003148)
        (ferret,1.001242)
        (fluidanimate,1.013916)
        (freqmine,1.016823)
        (raytrace,1.006587)
        (vips,1.417999)
        (x264,1.004860)
    };
    \addlegendentry{Maximum}
\end{axis}
\end{tikzpicture}
}
\caption{Energy overhead of PARSEC benchmark applications when run on an 8$\times$8 NoC with SeFaR. Full-system simulation has been run for the NoC with (a) one infected router in the NoC with one HT and (b) one infected router in the NoC with a maximum of four HTs in the NoC that affects four out of its five input ports.}
\end{figure}
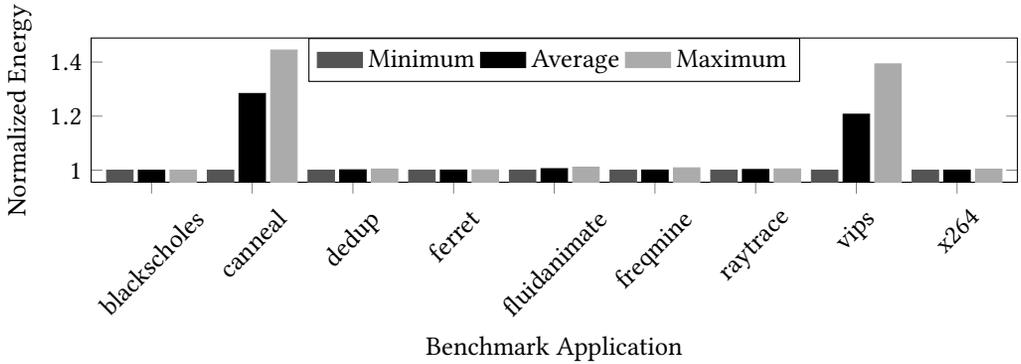
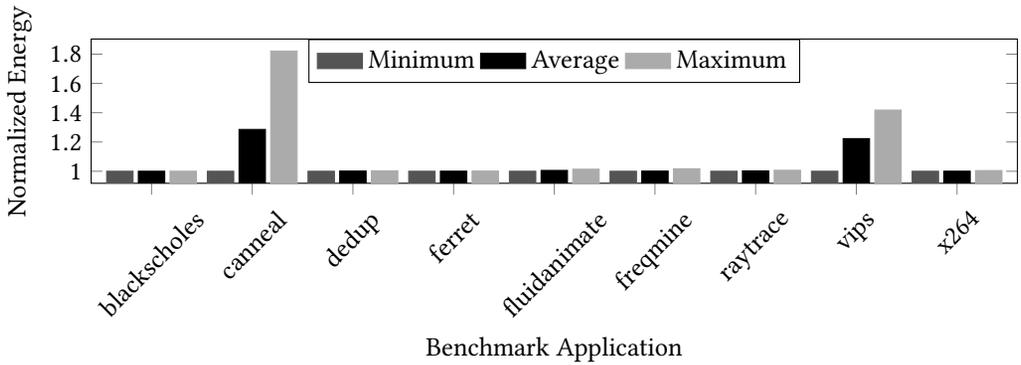

Similarly, Figs. \ref{fig:ENO1Port} and \ref{fig:ENO4Port} show a comparison of SeFaR with the baseline FT router regarding the overheads observed in the energy consumption while running the real benchmarks. The energy consumption overheads include the consumption overheads of the security blocks endowed in SeFaR. Fig. \ref{fig:ENO1Port} refers to the routers that are infected with a single HT, whereas Fig. \ref{fig:ENO4Port} refers to the routers that are infected with four HTs. For the case of a single HT in a packet drop router, the maximum overhead in the energy consumption has been observed for \textit{canneal}, which is 44.46\%, when compared with the baseline FT router. However, the average overhead in the energy consumption for each application does not exceed 28.36\% with respect to the baseline FT router, when a packet drop router is infected with a single HT. Similarly, for the case with four HTs in a packet drop router, the maximum overhead observed regarding the energy consumption for SeFaR has been 82.10\% for \textit{canneal}, when compared with the baseline FT router. Whereas, the average overhead in the execution time for this case does not exceed 28.58\%, for the considered real benchmarks. For the above set of evaluations, each benchmark application has been executed entirely, considering a link in the NoC to be faulty for the entire run of the application. However, if the link becomes defective while the application is being executed, the overheads observed would decrease depending on the temporal traffic distribution of the application under execution.

\subsubsection{Synthetic Traffic Patterns}
\label{subsub:synth}
\begin{figure}[!ht]
\centering
\subfloat[shuffle]{\resizebox{0.33\textwidth}{!}{
\begin{tikzpicture}
\begin{axis}[
ymax = 35,
legend pos = north west,
xlabel = PIR (Flit/Cycle/Node),
ylabel = APL (Cycle)
]
    \addplot[black, mark=diamond] plot table {Fault_free_shuffle.txt};
    \addlegendentry{Fault-free}
    \addplot[black, mark=*] plot table {Faulty5_NoHT_shuffle.txt};
    \addlegendentry{Faulty5\%-No HT}
    \addplot[black, mark=o] plot table {Faulty5_WithHT_shuffle.txt};
    \addlegendentry{Faulty5\%-HT}
    \addplot[black, mark=star] plot table {Faulty10_NoHT_shuffle.txt};
    \addlegendentry{Faulty10\%-No HT}
    \addplot[black, mark=oplus] plot table {Faulty10_WithHT_shuffle.txt};
    \addlegendentry{Faulty10\%-HT}
\end{axis}
\end{tikzpicture}
}}
\subfloat[transpose]{\resizebox{0.33\textwidth}{!}{
\begin{tikzpicture}
\begin{axis}[
ymax = 60,
legend pos = north west,
xlabel = PIR (Flit/Cycle/Node),
ylabel = APL (Cycle)
]
    \addplot[black, mark=diamond] plot table {Fault_free_transpose.txt};
    \addlegendentry{Fault-free}
    \addplot[black, mark=*] plot table {Faulty5_NoHT_transpose.txt};
    \addlegendentry{Faulty5\%-No HT}
    \addplot[black, mark=o] plot table {Faulty5_WithHT_transpose.txt};
    \addlegendentry{Faulty5\%-HT}
    \addplot[black, mark=star] plot table {Faulty10_NoHT_transpose.txt};
    \addlegendentry{Faulty10\%-No HT}
    \addplot[black, mark=oplus] plot table {Faulty10_WithHT_transpose.txt};
    \addlegendentry{Faulty10\%-HT}
\end{axis}
\end{tikzpicture}
}}
\subfloat[uniform]{\resizebox{0.33\textwidth}{!}{
\begin{tikzpicture}
\begin{axis}[
ymax = 45,
legend pos = north west,
xlabel = PIR (Flit/Cycle/Node),
ylabel = APL (Cycle),
]
    \addplot[black, mark=diamond] plot table {Fault_free_uniform.txt};
    \addlegendentry{Fault-free}
    \addplot[black, mark=*] plot table {Faulty5_NoHT_uniform.txt};
    \addlegendentry{Faulty5\%-No HT}
    \addplot[black, mark=o] plot table {Faulty5_WithHT_uniform.txt};
    \addlegendentry{Faulty5\%-HT}
    \addplot[black, mark=star] plot table {Faulty10_NoHT_uniform.txt};
    \addlegendentry{Faulty10\%-No HT}
    \addplot[black, mark=oplus] plot table {Faulty10_WithHT_uniform.txt};
    \addlegendentry{Faulty10\%-HT}
\end{axis}
\end{tikzpicture}
}}\\
\subfloat[shuffle]{\resizebox{0.33\textwidth}{!}{
\begin{tikzpicture}
\begin{axis}[
ymax = 4000,
legend pos = north west,
xlabel = PIR (Flit/Cycle/Node),
ylabel = PLP (W-Cycle)
]
    \addplot[black, mark=diamond] plot table {PLP_Fault_free_shuffle.txt};
    \addlegendentry{Fault-free}
    \addplot[black, mark=*] plot table {PLP_Faulty5_NoHT_shuffle.txt};
    \addlegendentry{Faulty5\%-No HT}
    \addplot[black, mark=o] plot table {PLP_Faulty5_WithHT_shuffle.txt};
    \addlegendentry{Faulty5\%-HT}
    \addplot[black, mark=star] plot table {PLP_Faulty10_NoHT_shuffle.txt};
    \addlegendentry{Faulty10\%-No HT}
    \addplot[black, mark=oplus] plot table {PLP_Faulty10_WithHT_shuffle.txt};
    \addlegendentry{Faulty10\%-HT}
\end{axis}
\end{tikzpicture}
}}
\subfloat[transpose]{\resizebox{0.33\textwidth}{!}{
\begin{tikzpicture}
\begin{axis}[
ymax = 11000,
legend pos = north west,
xlabel = PIR (Flit/Cycle/Node),
ylabel = PLP (W-Cycle)
]
    \addplot[black, mark=diamond] plot table {PLP_Fault_free_transpose.txt};
    \addlegendentry{Fault-free}
    \addplot[black, mark=*] plot table {PLP_Faulty5_NoHT_transpose.txt};
    \addlegendentry{Faulty5\%-No HT}
    \addplot[black, mark=o] plot table {PLP_Faulty5_WithHT_transpose.txt};
    \addlegendentry{Faulty5\%-HT}
    \addplot[black, mark=star] plot table {PLP_Faulty10_NoHT_transpose.txt};
    \addlegendentry{Faulty10\%-No HT}
    \addplot[black, mark=oplus] plot table {PLP_Faulty10_WithHT_transpose.txt};
    \addlegendentry{Faulty10\%-HT}
\end{axis}
\end{tikzpicture}
}}
\subfloat[uniform]{\resizebox{0.33\textwidth}{!}{
\begin{tikzpicture}
\begin{axis}[
ymax = 6000,
legend pos = north west,
xlabel = PIR (Flit/Cycle/Node),
ylabel = PLP (W-Cycle),
]
    \addplot[black, mark=diamond] plot table {PLP_Fault_free_uniform.txt};
    \addlegendentry{Fault-free}
    \addplot[black, mark=*] plot table {PLP_Faulty5_NoHT_uniform.txt};
    \addlegendentry{Faulty5\%-No HT}
    \addplot[black, mark=o] plot table {PLP_Faulty5_WithHT_uniform.txt};
    \addlegendentry{Faulty5\%-HT}
    \addplot[black, mark=star] plot table {PLP_Faulty10_NoHT_uniform.txt};
    \addlegendentry{Faulty10\%-No HT}
    \addplot[black, mark=oplus] plot table {PLP_Faulty10_WithHT_uniform.txt};
    \addlegendentry{Faulty10\%-HT}
\end{axis}
\end{tikzpicture}
}}
\caption{Average packet latency (APL) (a, b, and c) and power-latency product (PLP) (d, e, and f) versus packet injection rate (PIR) of different synthetic traffic patterns in an 8$\times$8 mesh NoC.}
\label{fig:aplplpsynovd}
\end{figure}
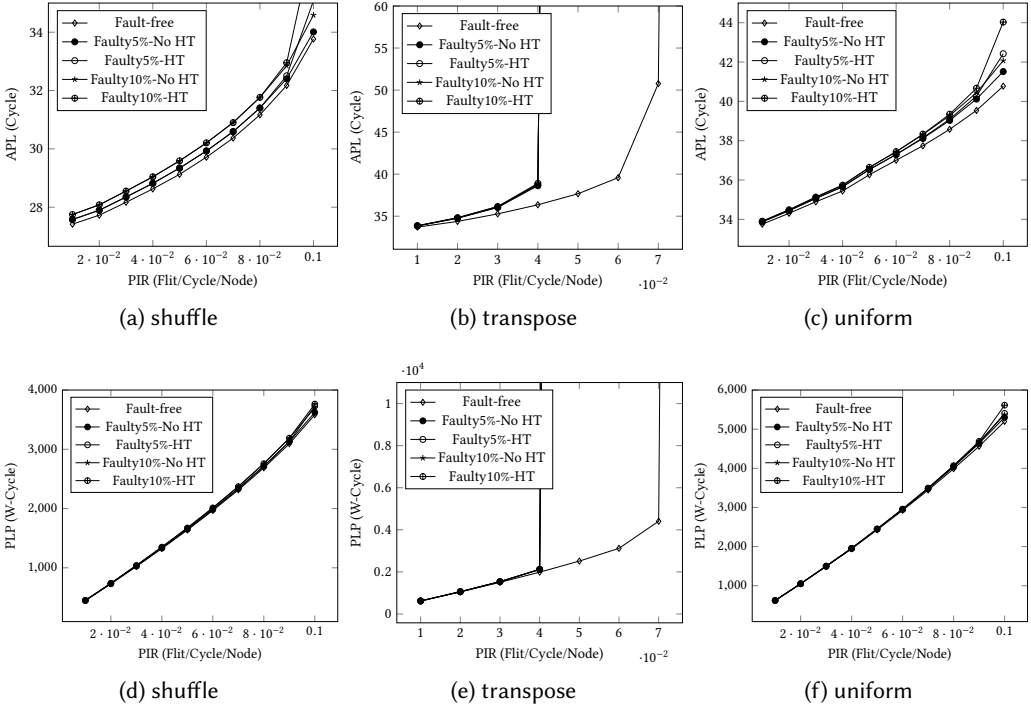

Fig. \ref{fig:aplplpsynovd} shows the plots of different configurations of an 8$\times$8 NoC endowed with SeFaR regarding the average packet latency (APL) and power-latency product (PLP) with varying packet injection rates (PIRs) while running the synthetic traffic patterns. Here, 'Fault-free' case refers to the NoC without any link fault. 'Faulty5\%-No HT' case refers to the NoC with 5\% link faults, but the HTs, which are infected in the routers that have these links at their output ports, are not triggered. 'Faulty5\%-HT' case refers to the NoC with 5\% link faults and all routers with those links serving as output links are infected with an HT. Similar is the convention for the NoC with 10\% link faults. From the plots, one may note that the APL and PLP of the NoC have similar trends for the \textit{shuffle} and \textit{uniform} traffic patterns, for all the five cases. This similarity is due to the nature of the spatial distribution of the traffic in the network. However, for the \textit{transpose} traffic pattern, the NoC with active HTs suffers a slight degradation in APL and PLP when compared to the NoC with link faults, but with dormant HTs. Though the plots in Fig. \ref{fig:aplplpsynovd} are shown for random locations of the faulty links, similar trends have been observed for all the five configurations of SeFaR when the locations of the defective links in the network change. From the plots, one may find that the overhead introduced in the NoC with faulty links and active HTs, because of the mitigation mechanism in the form of SeFaR, in the APL and PLP metrics are not so high when compared with the values observed when the NoC has faulty links, but dormant HTs.

\section{Design of a Comprehensive Secure Router}
\label{sec:comprouter}
    Several mitigation mechanisms proposed hitherto, which aim in restoring security in NoCs, primarily address one particular attack. This kind of attack-specific mitigation mechanism for NoCs may not be viable to be considered in practice, as each such architecture can nullify only a specific attack raised by the HTs. Thus, more generic mitigation mechanisms, which can mitigate multiple attacks, are necessary to extend the scope of Quality-of-Security Service (QoSS) in the future NoC architectures. One way of realizing a secure architecture of a router in the NoC is by leveraging the common properties of several mitigation mechanisms and integrating them into the router. This kind of hardening mechanism improves the immunity of the routers in the NoCs to resist and thwart several attacks.

    This section presents the architecture of one such router, namely comprehensive secure router (CSR), which can mitigate multiple attack scenarios that arise within the NoC routers. The architecture of CSR is a derived architecture, which has been derived from other secure architectures presented in the literature.
    
    \begin{figure}[!ht]
        \centering
        \includegraphics[width=0.4825\textwidth]{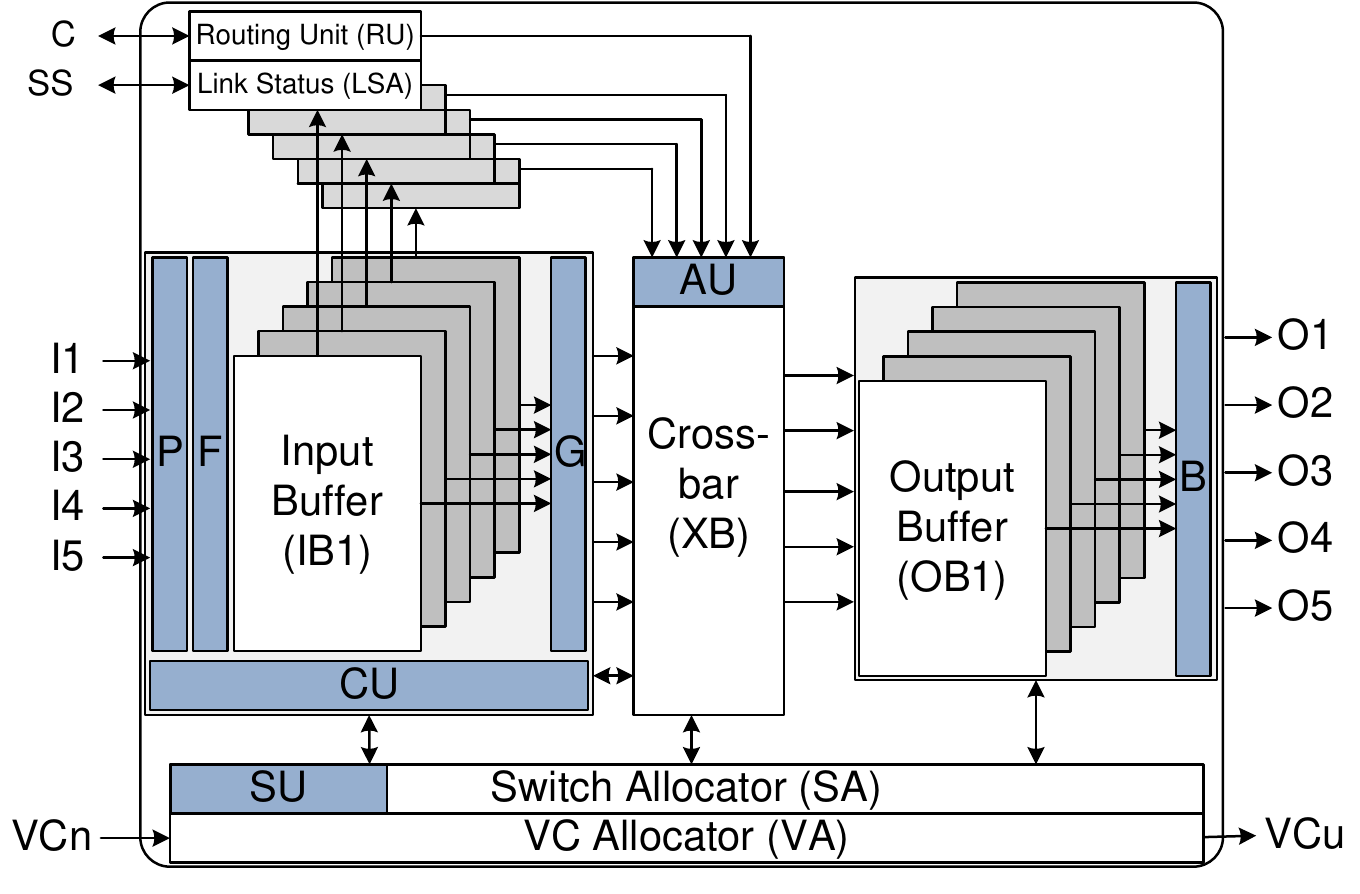}
        \caption{Microarchitecture of a secure router capable of mitigating several HT attacks.}
        \label{fig:comprouter}
    \end{figure}
    
    Fig. \ref{fig:comprouter} shows one such CSR, which can mitigate the attacks considered in \cite{Frey201715,ISCAS17}, along with the proposed packet drop attack. The blocks buffer shuffler (P), control unit (CU), and authentication unit (AU) help in mitigating the packet drop attack, whereas, the blocks security unit (SU) and buffer shuffler (P) help in mitigating the illegal packet request attack \cite{ISCAS17}. Similarly, the blocks F, G, and B conform to the blocks HTC1, HTC2, and HTC3, mentioned in \cite{Frey201715}, which help in mitigating the flit sabotage attacks that arise within the router.
    
    The design overhead of the router in Fig. \ref{fig:comprouter} compared to the baseline router mentioned in \cite{Frey201715}, which consists of 5 IO ports and 8 VCs/port with a flit-width of 32 bits, is found to be about 41\% in terms of area and 18\% in terms of power consumption.
    
\section{Conclusions and Future Work}
\label{sec:conclusions}
    Packet drop attack, a type of denial-of-service attack, has been proposed in the context of NoCs, in which packets are forced by the HTs to proceed to those output ports of a router that have faulty links. The HTs that can raise the proposed attack have been inserted inside the routers, which become active when an output link associated with such routers becomes faulty. Security modules, namely authentication unit, control unit, and buffer shuffler have been proposed to mitigate the anomaly raised because of the attacker HTs. Regarding overheads in performance, SeFaR suffers an average overhead of at most 0.8\% and 12.89\% in terms of execution time, when one and four active HTs are considered in the NoC while running real benchmarks. Similarly, regarding the overhead in energy consumption, SeFaR additionally consumes an average energy of at most 28.36\% and 28.58\%, when one and four active HTs are considered in the NoC while running real benchmarks. For synthetic traffic patterns, SeFaR with an active HT suffers from slight degradation in the average packet latency and power-latency product metrics, when compared with SeFaR without any HT. Area and power overheads of SeFaR, when put next to the baseline FT router, have been found to be 2.02\% and 0.90\%, respectively. Further, a comprehensive secure router, with area and power overheads of 41\% and 18\% compared to the baseline router, has been designed, which can mitigate multiple attack scenarios that arise within the routers of a NoC. Future work aims at improving the mitigation mechanism for the proposed attack.

\bibliographystyle{ACM-Reference-Format}
\bibliography{References.bib}

\end{document}